\documentclass[a4paper,superscriptaddress, nofootinbib]{revtex4}[11pt]
\pdfoutput=1

\usepackage{graphicx}% Include figure files
\graphicspath{{figs/}}
\usepackage {amsmath, amssymb, rotating}
\usepackage{hyperref}
\usepackage[usenames,dvipsnames]{color}
\usepackage{mdwlist} %to make compact lists: \begin{itemize*} ... \end{itemize*}
\usepackage{slashed}
\usepackage{soul}
\usepackage{xcolor}
\usepackage{verbatim}
\usepackage{wasysym}
\usepackage{cancel}
\usepackage{cleveref}

\definecolor{pyellow}{rgb}{0.99, 0.99, 0.59}
\definecolor{lgray}{rgb}{0.83, 0.83, 0.83}
\definecolor{bubbles}{rgb}{0.91, 1.0, 1.0}
\definecolor{airforceblue}{rgb}{0.36, 0.54, 0.66}
\definecolor{amaranth}{rgb}{0.9, 0.17, 0.31}
\definecolor{orangy}{rgb}{1.0, 0.49, 0.0}
\definecolor{flag}{rgb}{1.0, 0.03, 0.0}
\definecolor{bgrey}{rgb}{0.52, 0.52, 0.51}
\definecolor{cadmiumgreen}{rgb}{0.0, 0.42, 0.24}
\definecolor{applegreen}{rgb}{0.55, 0.71, 0.0}

\makeatletter
\def\@seccntformat#1{\csname the#1\endcsname.\quad} 

\renewcommand\section{\@startsection
  {section}{1}{0mm}%name, level, indent
  {-\baselineskip}%                beforeskip
  {0.5\baselineskip}%            afterskip
  {\normalfont\normalsize\bf}}%{\hspace{-.45cm}.\; #1}}% style

\renewcommand\subsection{\@startsection
  {subsection}{2}{0mm}%name, level, indent
  {-\baselineskip}%             beforeskip
  {0.5\baselineskip}%            afterskip
  {\normalfont\normalsize\bf}}%{\hspace{-.45cm}.\; #1}}% style

\renewcommand\paragraph{\@startsection
  {paragraph}{4}{\z@}%name, level, indent
  {1.25ex \@plus1ex \@minus.2ex}%     beforeskip
  {-1em}%            afterskip
  {\normalfont\normalsize\bfseries}}  
\makeatother

\begin{document}

\title{Reflections on Bubble Walls}

\author{Isabel Garcia Garcia}
\email{isabel@ias.edu}
\affiliation{Center for Cosmology and Particle Physics, Department of Physics, New York University, New York, NY 10003, USA}
\affiliation{School of Natural Sciences, Institute for Advanced Study, Princeton, NJ 08540, USA}

\author{Giacomo Koszegi}
\email{koszegi@physics.ucsb.edu}
\affiliation{Department of Physics, University of California, Santa Barbara, CA 93106, USA}

\author{Rudin Petrossian-Byrne}
\email{rpetross@ictp.it}
\affiliation{Abdus Salam International Centre for Theoretical Physics,
Strada Costiera 11, 34151, Trieste, Italy}

\begin{abstract} \begin{center}
	{\bf Abstract}\\
	
We discuss the dynamics of expanding bubble walls in the presence of massive dark photons whose mass changes as they cross the wall.
For sufficiently thin walls, we show that there exists a transient kinematic regime characterized by a constant reflection probability of longitudinal -- but not transverse -- modes.
This effect can have important implications for the dynamics of expanding vacuum bubbles in the early Universe.
Most notably, it leads to a new source of pressure on the expanding interface, featuring a  non-monotonic dependence on the $\gamma$-factor of the bubble walls and reaching a peak at intermediate $\gamma$-factors that we dub Maximum Dynamic Pressure.
When this pressure is large enough to halt the acceleration of the bubble walls, the difference in vacuum energy densities goes into making a fraction of the dark photons relativistic, turning them into dark radiation.
If the dark radiation remains relativistic until late times, an observable contribution to $\Delta N_\text{eff}$ is possible for phase transitions with strength $\alpha \sim 10^{-2} - 10^{-1}$.

\end{center} \end{abstract}

\maketitle

\begin{center} (Dated: \today) \end{center}

%\end{titlepage}

\tableofcontents

%\newpage

%\setcounter{equation}{0} \setcounter{footnote}{0}

\clearpage

%%%%%%%%%%%%%%%%%%%%%%%%%%%%%%%%%%%%%%%%%
\section{\label{sec:intro} \ Introduction}
%%%%%%%%%%%%%%%%%%%%%%%%%%%%%%%%%%%%%%%%%
Cosmological phase transitions in the early Universe that proceed via nucleation of bubbles are a well-motivated possibility in minimal extensions of the Standard Model, as well as in more general scenarios featuring hidden sectors with their own dynamics. Such first order transitions result in the emission of gravitational radiation~\cite{Witten:1984rs,Hogan:1986qda,Kosowsky:1991ua,Kosowsky:1992rz,Kamionkowski:1993fg} that current and future observatories may be able to detect in the form of a stochastic background of gravitational waves (see e.g.~\cite{Caprini:2015zlo,Caprini:2019egz} for reviews). An observation of this kind would provide unambiguous evidence for the existence of degrees of freedom beyond the Standard Model. This has spurred significant interest in the gravitational wave signatures of hidden sectors in recent times~\cite{Schwaller:2015tja,Breitbach:2018ddu,Helmboldt:2019pan,Jinno:2019jhi,Azatov:2019png,Nakai:2020oit,Halverson:2020xpg,Huang:2021rrk,Ertas:2021xeh,Dent:2022bcd,Fairbairn:2019xog,Jinno:2022fom}.

Despite the abundance of particle physics models susceptible of undergoing an out-of-equilibrium transition, our ability to make use of the resulting gravitational wave signal to extract information about the relevant dynamics is extremely limited. The most revealing features concern the frequency peak of the stochastic background, as well as its spectral shape at high frequencies. The former determines the epoch at which the transition takes place, whereas the latter contains information about the dominant source of gravitational radiation.
For example, if most of the energy released during the transition goes into accelerating the bubble walls (as in vacuum~\cite{Coleman:1977py}), these become relativistic and continue to expand at ever-increasing velocities. In this case, collisions of these ``run-away" bubbles constitute the main source of gravitational waves, and the resulting signal falls off as $f^{-1}$ at high frequencies~\cite{Huber:2008hg}.
Alternatively, pressure on the bubble walls due to particles in the thermal plasma may cause the expanding walls to reach a terminal speed.
In this case, most of the latent heat gets damped instead into the thermal fluid, and it is its subsequent motion that provides the dominant source of gravitational radiation~\cite{Espinosa:2010hh}. The high-frequency fall-off of the stochastic background is steeper, e.g.~decreasing as $f^{-4}$ for radiation sourced by sound waves~\cite{Hindmarsh:2015qta,Hindmarsh:2016lnk,Hindmarsh:2017gnf,Hindmarsh:2019phv}.
On the other hand, assuming radiation-domination at the time of gravitational wave production,
causal propagation restricts the low frequency shape of the spectrum to grow as $f^3$, independently of the dominant production channel~\cite{Caprini:2009fx}.~\footnote{More generally, the low-frequency shape of the stochastic background depends on the equation of state~\cite{Barenboim:2016mjm,Hook:2020phx} and on the existence of free-streaming particles~\cite{Weinberg:2003ur,Hook:2020phx}, and could thus provide non-trivial information about the Universe at early times.}

Understanding the dynamics of expanding bubble walls in the early Universe is clearly crucial to determine both quantitative and qualitative features of the resulting gravitational wave signal. But given the intrinsic degeneracy present in any stochastic background, it is equally important to explore alternative probes of the relevant dynamics.
For example, the upcoming LISA experiment will be sensitive to phase transitions at electroweak to multi-TeV scale temperatures, probing the nature of the electroweak phase transition and potentially shedding light on the dynamics behind baryogenesis and electroweak symmetry breaking. In this range of energies, the complementarity between LISA and current and future colliders will no doubt be key in furthering our understanding of physics at and around the weak scale (see e.g.~\cite{Morrissey:2012db,Caprini:2015zlo,Caprini:2019egz,AlAli:2021let} for reviews and references). Beyond (and below) the electroweak scale, phase transitions within hidden sectors may occur at virtually any temperature, and the corresponding stochastic background could fall anywhere from the low frequency range of PTA observatories \cite{NANOGRAV:2018hou,Lentati:2015qwp} ($10^{-9}-10^{-7} \ \text{Hz}$), to the high frequencies probed by LIGO ($10 - 10^3 \ \text{Hz}$), and beyond.
However, vacuum bubbles nucleated in hidden sector transitions may feature very different dynamics to those linked to the weak scale, and the relevant degrees of freedom may be inaccessible at laboratory experiments. Our work is motivated by the goal to more broadly understand the potential behavior of expanding bubbles in the early Universe, as well as to identify alternative predictions that may accompany an observable stochastic background of gravitational waves.

In a first order phase transition, bubbles of true vacuum are nucleated at rest, and begin to expand fueled by the difference in free energy densities at either side of the interface. In vacuum, the velocity of the bubble walls evolves according to~\cite{Coleman:1977py}
\begin{equation}
    \frac{d |\vec v|}{d t}
    = \frac{1}{\gamma^3 R_0}
    \propto \frac{1}{\gamma^3} \frac{\Delta V}{\sigma} \ ,
\label{eq:gamma_vacuum}
\end{equation}
with $\gamma = 1 / \sqrt{1 - \vec v^{\, 2}}$ the usual Lorentz $\gamma$-factor, $R_0 \propto \sigma / \Delta V$ the critical bubble radius, and where $\Delta V$ and $\sigma$ refer to the difference in vacuum energy densities and surface tension of the bubble wall.
In reality, bubbles do not expand against a sea of false vacuum, but rather within a non-trivial environment that in the early Universe must include the Standard Model plasma as well as, potentially, other ingredients such as dark matter.
As the bubble wall speed grows, friction from the surrounding environment can exert a pressure on the interface that opposes the expansion of the bubble walls.
Their evolution can still be written as in \cref{eq:gamma_vacuum}, after the replacement \cite{Arnold:1993wc,Moore:1995si}:
\begin{equation}
   \Delta V \rightarrow \Delta V - \mathcal{P} \ ,
\end{equation}
where $\mathcal{P}$ refers to the frictional pressure resulting from the interaction between the interface and the surrounding medium, and is in general dependent on the speed of the bubble wall.

Two qualitatively different scenarios are thus possible. If $\Delta V \gg \mathcal{P}$ during the entire evolution of the bubble walls, these effectively behave as if they were in vacuum and most of the energy released as the bubbles grow goes into accelerating the expanding interface. The walls then ``run away" -- that is, they continue to expand with ever increasing velocities.
Alternatively, if $\mathcal{P}$ grows large enough so as to neutralize the difference in vacuum energies, $\Delta V = \mathcal{P}$, the bubble walls reach an equilibrium regime of constant speed. Once in equilibrium, the fraction of the total energy that becomes localized on the interface quickly becomes tiny.
Instead, most of the energy gets damped into whatever sector of the surrounding environment is responsible for halting the acceleration of the bubble walls.
Calculating the pressure experienced by bubble walls as they expand is a classic problem~\cite{Turok:1992jp,Dine:1992wr,Liu:1992tn,Arnold:1993wc,Moore:1995ua,Moore:1995si} that has received renewed attention in recent times~\cite{Bodeker:2009qy,Espinosa:2010hh,Leitao:2015ola,Bodeker:2017cim,Azatov:2020ufh,Ai:2021kak,Mancha_2021,Bea:2021zsu,Gouttenoire:2021kjv,Azatov:2022tii,Laurent:2022jrs,DeCurtis:2022hlx}.

Within a thermal plasma, particles with phase-dependent mass create a pressure on the expanding walls that asymptotes to a constant $\mathcal{P}_\infty \sim \Delta m^2 T^2$ in the ultra-relativistic limit
\footnote{The ultra-relativistic limit refers to the kinematic regime where the energy of the incident particles in the rest frame of the bubble wall is the largest energy scale. Alternatively, this corresponds to the limit $\gamma \rightarrow \infty$ for the $\gamma$-factor of the bubble wall.},
independently of the type of particle~\cite{Bodeker:2009qy}.
An additional source of friction may be present if the spectrum contains gauge bosons, of the form $\mathcal{P}_\infty \sim \gamma g^2 \Delta m_v T^3$, with $g$ the relevant gauge coupling and $\Delta m_v$ the change in the mass of the gauge boson at either side of the bubble wall~\cite{Bodeker:2017cim,Azatov:2020ufh,Gouttenoire:2021kjv}. This effect has its origin in the transition radiation emitted by charged particles as they cross the wall and its $\gamma$-dependence can easily render it the most significant source of friction on fast expanding bubbles. Indeed, if the electroweak phase transition were first order, transition radiation would likely cause the bubble walls to reach an equilibrium $\gamma$-factor as low as $\gamma_\text{eq} = \mathcal{O}(10)$~\cite{Bodeker:2017cim}.

A crucial aspect of all sources of friction known so far is that the corresponding pressure is a monotonically increasing function of the wall speed.
As put forward in \cite{Bodeker:2009qy,Bodeker:2017cim}, this allows for a simple criterion to determine whether bubble walls during a cosmological phase transition become run-away, by comparing the pressure in the ultra-relativistic limit, $\mathcal{P}_\infty$, to the difference in vacuum energy density across the wall: 
\begin{equation}
    \text{Run-away criterion:}
    \qquad \qquad
    \Delta V > \mathcal{P}_\infty
    \qquad \qquad
    \text{\cite{Bodeker:2009qy,Bodeker:2017cim}} \ .
\label{eq:runaway_naive}
\end{equation}

In this article, we discuss a new physical effect that can qualitatively alter the dynamics of bubble walls during a cosmological phase transition. Namely, the existence of a transient relativistic regime characterized by an approximately constant reflection probability of longitudinal massive vectors off an expanding interface. Effectively, the wall behaves temporarily like an imperfect mirror that reflects a fraction of longitudinal -- but not transverse -- modes. Two conditions need to be satisfied for this regime to be accessible: (\emph{i}) that the expansion of the bubble walls takes place against a population of massive vectors whose mass changes across the interface; and (\emph{ii}) that the expanding walls are sufficiently ``thin". By thin, we mean that the wall thickness (in the rest frame of the bubble wall) be much smaller than the Compton wavelength of the massive vector,~i.e.~
\begin{equation}
	L \ll m^{-1} \ .
\label{eq:cond_ir}
\end{equation}
In this case, the regime of constant longitudinal reflection corresponds to Lorentz $\gamma$-factors in the range
\begin{equation}
	1 \ll \gamma \ll (L m)^{-1} \ ,
\label{eq:def_ir}
\end{equation}
ending when $\gamma$ is so large that the Lorentz-contracted Compton wavelength of the dark photon becomes smaller than the wall thickness.
We will refer to this kinematic regime as the region of ``inter-relativistic" motion.
Eq.(\ref{eq:cond_ir}) ensures that $(L m)^{-1} \gg 1$, and that this regime is indeed accessible during the evolution of the expanding bubbles.
Once $\gamma$ becomes $\gg (L m)^{-1}$, reflection probabilities for all polarizations die off exponentially -- a well-known feature of the ultra-relativistic limit.

Most notably, the effect described above leads to an additional source of friction on expanding bubble~walls.
Unlike previously known cases, the corresponding pressure features a characteristic non-monotonic dependence on the relevant $\gamma$-factor, reaching a maximum at $\gamma \sim (L m)^{-1}$ before turning-off at larger values. In (superficial) analogy with the behavior of spacecraft shortly after launch, we will refer to this pressure peak as Maximum Dynamic Pressure.
Its existence can make it much harder for bubble walls to become run-away than previously believed, and  we will show that \cref{eq:runaway_naive} can be qualitatively misleading in phase transitions where the bubble walls expand against an existing population of phase-dependent massive dark photons.

This article is dedicated to deriving the claims made in the previous two paragraphs, as well as illustrating some of their phenomenological consequences.
We will refer to the temperature of the Standard Model plasma at the epoch of the phase transition as $T_*$, and denote the phase transition strength via the usual dimensionless quantity \cite{Caprini:2015zlo,Caprini:2019egz}:
\begin{equation} \label{eq:alpha_def}
    \alpha  \equiv \frac{\Delta V}{\rho_\text{SM} (T_*)} = \frac{\Delta V}{\frac{\pi^2}{30} g_*(T_*) T_*^4} \ .
\end{equation}
We will focus on bubble walls that expand against a population of cold and non-interacting dark photons. Despite its simplicity, this system will be relevant in some physically interesting cases, such as when the dark photons furnish the dark matter \cite{Hu:2000ke,Hui:2016ltb,Nelson:2011sf,Arias:2012az,Graham:2015rva,Agrawal:2018vin} -- a well-motivated benchmark that we often refer to throughout this work.
When the bubble walls reach an equilibrium regime as a result of longitudinal reflections, the fraction of the total energy that goes into accelerating the bubble walls becomes increasingly small. Instead, most of the available energy goes into making the reflected dark photons relativistic, turning them into dark radiation.
If the dark radiation remains relativistic until late times, an observable contribution to $\Delta N_\text{eff}$ is possible. In particular, current bounds on $\Delta N_\text{eff}$ could probe phase transitions with strength $\alpha \gtrsim 10^{-1}$, whereas CMB S-4 measurements could be sensitive to scenarios down to $\alpha \sim 10^{-2}$ for all relevant frequencies.

Extensions of the Standard Model featuring massive vectors are popular both because of their minimality as well as their potential to furnish the dark matter, with a variety of production mechanisms spanning a wide mass range~\cite{Nelson:2011sf,Arias:2012az,Graham:2015rva,Agrawal:2018vin}.
Before we move on, let us summarize why the existence of \emph{massive} dark photons whose mass changes in the course of our cosmological history is not only a well-motivated possibility, but may be an unavoidable feature in a wide class of models.
At the renormalizable level, the physical system that we focus on is described by a lagrangian of the form
\begin{equation}
    \mathcal{L} \supset \frac{1}{2} \left( \partial_\mu \phi \right)^2 - V(\phi) - \frac{1}{4} F_{\mu \nu} F^{\mu \nu} + \frac{1}{2} m^2 V_\mu V^\mu \ ,
\label{eq:L_dim4}
\end{equation}
where $\phi$ is a real scalar field and $V_\mu$ is the massive dark photon.
By assumption, the potential for $\phi$ features two non-degenerate vacua such that the scalar sector undergoes a phase transition in the early Universe.
WLOG we take the false and the true vacuum to lie at $\langle \phi \rangle = 0$ and $\langle \phi \rangle = v$.
$V(\phi)$ may be a finite temperature effective potential, or it may be a zero-temperature potential in which case the transition proceeds via quantum tunneling.
If the scalar sector is in thermal equilibrium with the Standard Model then $T_* \sim v$, whereas $T_* \ll v$ is possible if the transition is `super-cooled', or if $\phi$ belongs in a hidden sector that is decoupled from the thermal plasma.
At nucleation, the thickness of the bubble walls is determined by the features of the scalar potential, and typically $L \sim (\sqrt{\lambda} \, v)^{-1}$, with $\lambda$ the typical quartic coupling in $V(\phi)$. The size of the dimensionless combination $T_*  L$ will have a quantitative effect on our results, as we will discuss when relevant.

At the level of \cref{eq:L_dim4}, the scalar and dark photon sectors are fully decoupled. Beyond the renormalizable terms of \cref{eq:L_dim4}, this need not remain true. For example, the following operator
\begin{equation}
    \mathcal{L} \supset \frac{\kappa}{2} \phi^2 V^\mu V_\mu \ ,
\label{eq:model_EFT_intro}
\end{equation}
leads to an additional contribution to the vector mass in the true vacuum, of the form $\Delta m^2 = \kappa v^2$. As the scalar vev `turns on' in the early Universe the dark photon mass will shift accordingly. Of course, the scalar field may be complex instead of real, which can be trivially accommodated by writing $|\phi|^2$ instead. Indeed, an effective interaction $\propto |H|^2 V^\mu V_\mu$, with $H$ the Standard Model Higgs doublet, would lead to a shift in the dark photon mass before and after the electroweak phase transition.
We emphasize that \cref{eq:model_EFT_intro} cannot be forbidden on the basis of symmetry, and its manifest lack of gauge redundancy is a moot point given that the theory under consideration already contains a mass for $V^\mu$.
Legalistically, one might object to \cref{eq:model_EFT_intro} on the basis that it differs from a St\"uckelberg mass, $\frac{1}{2} m^2 V^\mu V_\mu$, in that the former is not a renormalizable interaction and demands UV-completion. However, upholding the laws of effective field theory, we have no choice but to overrule this objection.
Accepting the existence of a finite cutoff in our description of nature, a theory with a massive dark photon and scalar degrees of freedom will in general feature effective interactions as in \cref{eq:model_EFT_intro}.

Indeed, a non-zero $\kappa$ in \cref{eq:model_EFT_intro} sets an upper bound on the scale of UV completion.
Provided that the term in \cref{eq:model_EFT_intro} only accounts for a subleading contribution to the overall mass of the dark photon in the true vacuum ($\Delta m^2 \ll m^2$), the upper bound on the UV cutoff can be conveniently written as \cite{Lebedev:2011iq}
\begin{equation}
    \Lambda \lesssim \frac{4 \pi v}{\sqrt{\Delta m^2 / m^2}} \ .
\label{eq:cutoff_intro}
\end{equation}
Here, we indeed focus on cases where the change in the dark photon mass is tiny.
This allows for a separation between the scale of the phase transition, $v$, and the UV cutoff, allowing us to neglect the effect of heavy degrees of freedom on the dynamics of the phase transition and focus instead on the consequences of non-zero $\Delta m^2$ on the evolution of the bubble walls.
As we will see, even when $\Delta m^2 / m^2 \ll 1$, the implications for the evolution of cosmological vacuum bubbles can be significant.
\footnote{Although most of our subsequent discussion will proceed within the effective theory defined in \cref{eq:L_dim4}-\ref{eq:model_EFT_intro}, we discuss in \cref{sec:model} how this effective description can arise from a UV-complete model.}

An important comment before we proceed. As made clear in the preceding paragraph, the content of this paper is only relevant in the presence of vector bosons that are \emph{massive} at either side of the bubble wall.
Our results therefore do not affect the pressure created by \emph{massless} gauge bosons that gain a mass as they cross the wall, and so we have nothing to add to e.g.~the pressure created by $W$ and $Z$ bosons during a first order electroweak phase transition.

The rest of this paper is organized as follows. We begin in \cref{sec:setup} with a summary of the physical set-up that we will focus on in the remainder of this manuscript.
%%%%%%%%%%%%%%%
In \cref{sec:probs}, we present our calculation of the reflection probability for phase-dependent massive vectors.
Fig.~\ref{fig:coeffs} illustrates the main result of this section: the existence of a transient relativistic regime characterized by an approximately constant reflection probability of longitudinally polarized dark photons.
%%%%%%%%%%%%%%%
Section~\ref{sec:friction} focuses on fleshing out the consequences of our results for the evolution of bubble walls in the early Universe, including the existence of a Maximum Dynamic Pressure in~\cref{sec:mdp}, as well as a self-consistent determination of the equilibrium $\gamma$-factor when this pressure is large enough to halt the acceleration of the bubble walls in \cref{sec:gamma_eq}.
%%%%%%%%%%%%%%%
The fate of the reflected dark photons depends sensitively on a variety of considerations, most notably on whether the sector undergoing the phase transition is hot or cold, as we discuss in \cref{sec:fate_dr}.
%%%%%%%%%%%%%%%
We summarize our conclusions in section \ref{sec:conclusions}, and a number of appendices supplement the discussion in the main text.

%%%%%%%%%%%%%%%%%%%%%%%%%%%%%%%%%%%%%%%%%
\section{\label{sec:setup} \ Set-up}
%%%%%%%%%%%%%%%%%%%%%%%%%%%%%%%%%%%%%%%%%

The rest of this article is dedicated to calculating the pressure on an expanding bubble wall due to an existing population of phase-dependent massive dark photons and to discussing the implications of our results.
With this goal, we consider an expanding planar interface, representing a portion of a sufficiently large bubble wall, moving with local velocity $\vec v$ and corresponding $\gamma$-factor $\gamma \equiv 1 / \sqrt{1 - \vec v^{\, 2}}$. The wall is not expanding in vacuum, but rather against a population of cold and non-interacting massive vector bosons with number density $n_V$.
WLOG, we take the velocity of the bubble wall in the rest frame of the dark photons to be along the negative $z$-axis, $\vec v = - |\vec v| \hat z$.
At leading order, to compute the pressure on the wall we need the momentum transfer from particles that either reflect or transmit across the interface. Our assumption that the dark photon sector is non-interacting allows us to consider the individual interactions of particles with the wall, although this approach is more generally valid whenever the relevant mean free path exceeds the wall thickness.

\begin{figure}[h]
  \centering
  \includegraphics[scale=0.9]{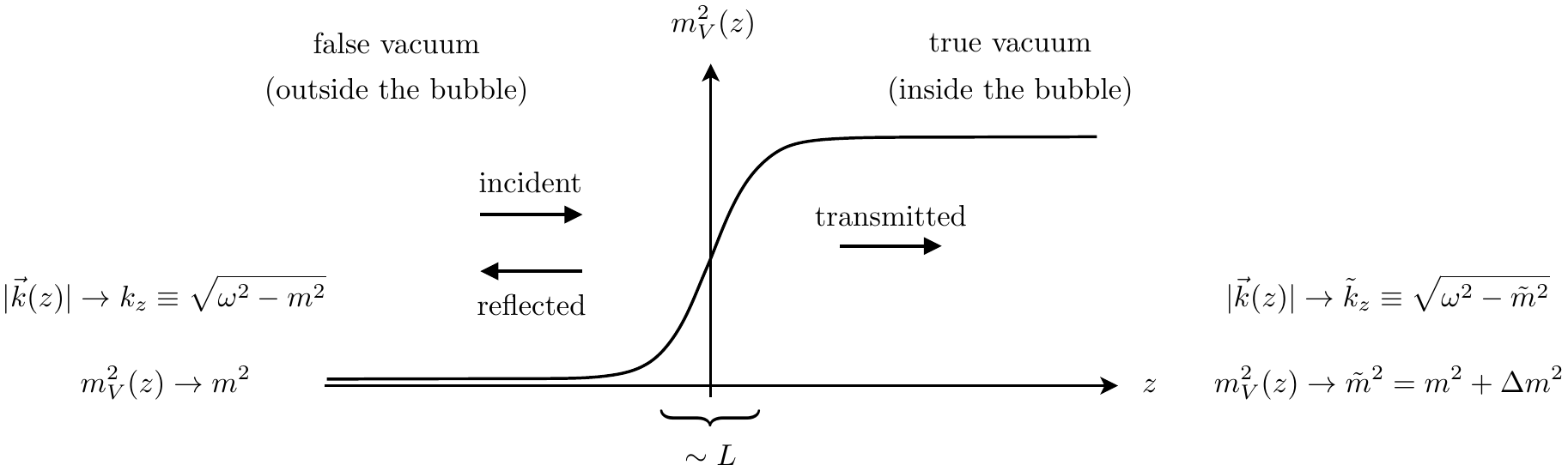}
  \caption{In the rest frame of the bubble wall, a dark photon wind hits the interface from the region of false vacuum. The energy of the incident dark photons, $\omega = \gamma m$, remains conserved in the interaction with the wall, whereas momentum along the $z$-direction changes resulting in a net momentum transfer to the interface. $L$ refers to the  wall thickness, i.e.~the typical length scale over which the order parameter varies. In this article, we focus on the case where the vector mass changes by a small amount at either side of the bubble wall, i.e.~$\Delta m^2 \ll m^2 \simeq \tilde m^2$. Much of our work will be concerned with the dynamics of the inter-relativistic kinematic regime of \cref{eq:def_ir} where $m \ll \omega \ll L^{-1}$.}
  \label{fig:setup}
\end{figure}

Like previous work focused on computing pressure on expanding bubbles, we find it convenient to work in the rest frame of the interface. In this frame, the scalar order parameter characterizing the transition varies only as a function of the spatial coordinates, which in our convention will be along the $z$-axis. A dark photon wind moving from the false to the true vacuum hits the bubble wall with velocity $- \vec v = |\vec v| \hat z$. All particles hit the interface at normal incidence and there are no particles traveling in the opposite direction -- both observations follow from our assumption that the dark photon sector is cold
\footnote{Quantitatively, the assumption of normal incidence requires that, in the wall frame, the normal component of the dark photon momentum is much greater than the components along the plane of the bubble wall, i.e.~$|\vec k_\perp| \gg |\vec k_\parallel|$. For a gas of dark photons at temperature $T_\text{dp} \ll m$ and a relativistic wall, $|\vec k_\perp| \sim \gamma m \gg |\vec k_\parallel| \sim \sqrt{T_\text{dr} m}$ provided $\gamma \gg \sqrt{T_\text{dp} / m}$, which is satisfied trivially. It should be be straightforward to generalize our results beyond these assumptions.}.
This set-up is depicted in~\cref{fig:setup}.
%%%%%%%%%%%
%%%%%%%%%%%

By assumption, the dark photon mass depends on the order parameter characterizing the phase transition. As anticipated in the Introduction, and further discussed in \cref{sec:model}, realistic situations often feature a dependence of the form $m^2_V (z) = m^2 + \kappa v (z)^2$, with $v$ the relevant order parameter and $\kappa$ a constant determined by the underlying model.
However, for most of our discussion, it will be sufficient to just assume that the vector mass-squared parameter is $z$-dependent, with asymptotic values $m^2$ and $\tilde{m}^2$ for large negative and positive $z$, as indicated in~\cref{fig:setup}. We define the difference in squared masses as $\Delta m^2 \equiv \tilde{m}^2-m^2>0$ and we will always assume that $\Delta m^2 \ll m^2 \simeq \tilde m^2$, in keeping with the discussion around~\cref{eq:cutoff_intro}.

Since the wall background is independent of $t$, the energy of the incident particles, $\omega = \gamma m$, remains conserved in the interaction with the wall, whereas momentum in the $z$-direction changes in a way that results in a net momentum transfer to the interface. In total, the pressure exerted on the wall as a result of this dark photon wind can be written as
\begin{equation}
	\mathcal{P} = \gamma | \vec v | n_V \, \times \frac{1}{3} \, \sum_\lambda \left( R_\lambda \, \Delta k_R + T_\lambda \, \Delta k_T \right) \ .
\label{eq:Pgamma}
\end{equation}
The factor $\gamma | \vec v | n_V$ corresponds to the flux of incoming massive vectors hitting the wall  (in the wall frame), whereas the quantity in parentheses represents the average momentum transfer to the wall from an incoming particle with fixed polarization $\lambda$, with $R_\lambda$ and $T_\lambda$ the corresponding reflection and transmission probabilities.
The momentum transfer from reflected and transmitted particles is given by $\Delta k_R = 2 k_z$ and $\Delta k_T = k_z - \tilde{k}_z$, with $k_z$ and $\tilde k_z$ the asymptotic transverse momenta at either side of the bubble wall, as defined in \cref{fig:setup}. The sum over $\lambda$ includes the three physical polarizations of a massive vector, and the lack of mixing between different polarizations in \cref{eq:Pgamma} follows as a result of normal incidence.

For example, in the regime of ultra-relativistic motion -- when $\omega$ is the largest energy scale in the problem, $\omega \gg m, L^{-1}$ -- reflection and transmission probabilities must asymptote to zero and unity respectively, and \cref{eq:Pgamma} takes the form,
\begin{equation}
	\mathcal{P}_\infty = \lim_{\gamma \rightarrow \infty} \mathcal{P} \simeq \gamma n_V \Delta k_T = \rho_V \frac{\Delta m^2}{2m^2} \ ,
\label{eq:P_UR}
\end{equation}
where we have used $\Delta k_T \simeq \Delta m^2 / (2 \gamma m)$, and $\rho_V = m n_V$ is the energy density of dark photons.
If the massive vector population outside of the bubble wall was instead described in terms of a fully thermal distribution, with $T \gg m$, we would have $\omega \sim \gamma T$ and $n_V \sim T^3$, and therefore $\mathcal{P}_\infty \sim T^2 \Delta m^2$ as is well known~\cite{Bodeker:2009qy}. As seen in \cref{eq:P_UR}, the frictional pressure reaches a constant (independent of the wall velocity) in this limit.

A comment is in order before we move on. Comparing the asymptotic pressure in \cref{eq:P_UR} to the difference in vacuum energy densities, one finds
\begin{equation}
\label{eq:dm_neglect}
    \frac{\mathcal{P}_\infty}{\Delta V} \
    \simeq \ \frac{\Delta m^2}{2 m^2} \ \frac{\rho_V}{\Delta V} \
    \sim \ \alpha \ \frac{\Delta m^2}{m^2} \ \frac{\rho_V (T_*)}{\rho_\text{dm} (T_*)} \ \frac{\rho_\text{dm} (T_*)}{\rho_\text{SM} (T_*)} \ ,
\end{equation}
where $\alpha$ is defined in \cref{eq:alpha_def} and in the last step we have made it explicit that all the relevant energy densities are to be evaluated at the phase transition epoch. The ratio $\rho_V / \rho_\text{dm}$ in \cref{eq:dm_neglect} is 1 if the dark photons account for all of the dark matter, whereas $\alpha$ and $\Delta m^2 / m^2$ are $\ll 1$ as discussed previously. The ratio of dark matter to Standard Model radiation in the early Universe is $\rho_\text{dm} / \rho_\text{SM} \lll 1$, making \cref{eq:dm_neglect} correspondingly tiny. As per the run-away criterion of \cref{eq:runaway_naive}, one would be tempted to conclude that a small change in the dark matter mass would have a negligible effect on the evolution of cosmological bubble walls.
Our results will show that -- in general -- this expectation can be mistaken.

To proceed, we must obtain reflection and transmission probabilities, as a function of $\gamma$, for massive vectors interacting with the non-trivial background of the bubble wall. This problem reduces to solving the equations of motion for massive electromagnetism with a spatially varying photon mass, as we discuss next.

%%%%%%%%%%%%%%%%%%%%%%%%%%%%%%%%%%%%%%%%%
\section{\label{sec:probs} \ Reflection and transmission probabilities}
%%%%%%%%%%%%%%%%%%%%%%%%%%%%%%%%%%%%%%%%%

We begin in \cref{sec:massiveEM} with a brief discussion regarding massive electromagnetism with a spatially varying vector mass.
In \ref{sec:step}, we obtain reflection and transmission probabilities in one the of the few non-trivial cases where an analytic solution is accessible: a step function change in the mass of the dark photon. This corresponds to the limit of vanishing wall thickness, and it provides an accurate description of the system in the regime $\omega \ll L^{-1}$. Considering a step wall allows us to illustrate one of our main results: that the reflection probability for longitudinal modes asymptotes to a constant in the inter-relativistic regime of \cref{eq:def_ir}.
In \ref{sec:smooth} we turn to the realistic situation of finite width, and show that the existence of a relativistic regime of near-constant longitudinal reflection is a generic feature of thin walls with finite thickness.

%%%%%%%%%%%%%%%%%%%%%%%%%%%%%%%%%%%%%%%%%
\subsection{\label{sec:massiveEM} \ Massive electromagnetism}

In the absence of charged sources, Maxwell's equations in the presence of a varying vector mass read
\begin{equation}
	\partial_\mu F^{\mu \nu} + m^2_V (x) V^\nu = 0 \ ,
\label{eq:EOM}
\end{equation}
which, in turn, imply the consistency condition
\begin{equation}
	\partial_\mu \left( m^2_V (x) V^\mu \right) = 0 \ .
\label{eq:condition}
\end{equation}
As is well known, when the vector mass is constant, $m_V^2(x) = m^2$, \cref{eq:condition} reduces to the familiar requirement that $\partial_\mu V^\mu = 0$, and Maxwell's equations admit plane-wave solutions of the form
\begin{equation}
	V^\mu (x) = v^\mu e^{-i k \cdot x} \qquad \qquad \text{(constant vector mass)} \ ,
\label{eq:Amu_constant}
\end{equation}
for all $k^\mu$ such that $k^2 = m^2$, and the $v^\mu$ are momentum-dependent complex coefficients satisfying $k_\mu v^\mu= 0$.~\footnote{The field $V^\mu$ is of course real-valued. It is implicitly assumed that only the real part of \cref{eq:Amu_constant} has physical significance. The same applies to all other expressions of this form in the remainder of this paper.} As usual, the $v^\mu$ can be written as a sum over unit-normalized polarization vectors, one for each of the three physical degrees of freedom of a massive spin-1 particle. For $\vec k = \pm | \vec k | \hat z$, a convenient basis is given by
\begin{equation}
	\varepsilon^\mu_x = (0,1,0,0) \ , \qquad \varepsilon^\mu_y = (0,0,1,0)
	\qquad \text{and} \qquad
	\varepsilon^\mu_l = \left( \frac{ | \vec k | }{m} , 0 , 0 , \pm \frac{\omega}{m} \right) \ .
\end{equation}

When the photon mass has a non-trivial profile, analytic solutions to \cref{eq:EOM}-(\ref{eq:condition}) only exist in some special cases. Assuming the vector mass-squared features no time dependence, as appropriate in the rest frame of the bubble wall, \cref{eq:condition} can be written as
\begin{align}
	\partial_\mu V^\mu 	= - \left( \vec \nabla \log m_V^2 (\vec x) \right) \cdot \vec V \ .
\label{eq:condition_z}
\end{align}
If $\vec k \parallel \vec \nabla m_V^2$, as in the case of normal incidence, \cref{eq:condition_z} is non-vanishing for the longitudinal component, whereas $\partial_\mu V^\mu = 0$ for the transverse modes. As summarized around \cref{fig:setup}, this is indeed the set-up that we focus on in this work.

%%%%%%%%%%%%%%%%%%%%%%%%%%%%%%%%%%%%%%%%%
\subsection{\label{sec:step} \ A step wall}

We will first consider a step-function change in the vector mass:
\begin{equation}
    m_V^2 (z) = m^2 + \Delta m_V^2(z) \qquad \text{with} \qquad  \Delta m_V^2 (z)=  \Delta m^2 \, \Theta (z) \ ,
\label{eq:mass_step}
\end{equation}
where $\Delta m^2 = \tilde{m}^2 - m^2 > 0$ and $\Theta(z)$ is the Heaviside step function. With this mass profile, the equations of motion feature plane-wave solutions on either side of the wall localized at $z=0$. These can be written as
\begin{equation}
	V^\mu_\perp (t, z) = 	e^{-i \omega t}
					\begin{cases} (0, 1, 1, 0) e^{i k_z z} + r_\perp (0, 1, 1, 0) e^{- i k_z z} & \quad z<0 \\
					t_\perp (0, 1, 1, 0) e^{i \tilde{k}_z z} & \quad z>0
					\end{cases} 
\label{eq:Aperp}
\end{equation}
for the transverse modes, \footnote{We have taken advantage of rotational invariance in the $x-y$ plane to set $t_x = t_y \equiv t_\perp$ and $r_x = r_y \equiv r_\perp$.} and
\begin{equation}
	V^\mu_l (t, z) = 	e^{-i \omega t}
					\begin{cases} \left( \frac{k_z}{m}, 0,0, \frac{\omega}{m} \right) e^{i k_z z} + r_l \left( \frac{k_z}{m} ,0,0, - \frac{\omega}{m} \right) e^{- i k_z z} & \quad z<0 \\
					t_l \left( \frac{\tilde{k}_z}{\tilde{m}} , 0,0, \frac{\omega}{\tilde{m}} \right) e^{i \tilde{k}_z z} & \quad z>0
					\end{cases} 
\label{eq:Al}
\end{equation}
for the longitudinal component, with $k_z = \sqrt{\omega^2 - m^2}$ and $\tilde k_z = \sqrt{\omega^2 - \tilde m^2}$ as defined in \cref{fig:setup}.
The overall normalization of \cref{eq:Aperp}-(\ref{eq:Al}) is arbitrary, and with this choice the reflection and transmission probabilities are given by
\begin{equation}
	R_\alpha = |r_\alpha|^2 \qquad \text{and} \qquad T_\alpha = \frac{\tilde{k}_z}{k_z} |t_\alpha |^2 \ ,
\end{equation}
with $\alpha = \perp, l$ for transverse and longitudinal modes respectively.

We can now obtain analytic solutions for both reflection and transmission probabilities by integrating the equations of motion across the interface. Let us discuss the transverse modes first. As discussed below \cref{eq:condition_z}, the transverse modes satisfy $\partial_\mu V^\mu_\perp = 0$, and \cref{eq:EOM} reads $(\Box + m_V^2 (z)) V_\perp^\mu = 0$. We then have:
\begin{equation}
	\lim_{\epsilon \rightarrow 0} \int_{- \epsilon}^{+ \epsilon} dz \, \left( \Box + m_V^2 (z) \right) V_\perp^\mu = 0
	\qquad \Rightarrow \qquad
	\partial_z V^\mu_\perp \ \text{is continuous at} \ z = 0 \ .
\end{equation}
Combined with the requirement that $V^\mu_\perp$ itself remains continuous, we can solve for $r_\perp$ and $t_\perp$. In particular, the reflection probability is given by
\begin{equation}
	R_\perp = |r_\perp|^2 = \left| \frac{k_z - \tilde{k}_z}{k_z + \tilde{k}_z} \right|^2 \xrightarrow{\omega \gg m, \tilde{m}} \left( \frac{\Delta m^2}{4 \omega^2} \right)^2 = \gamma^{- 4} \left( \frac{\Delta m^2}{4 m^2} \right)^2 \ .
\label{eq:Rperp}
\end{equation}
Unsurprisingly, $R_\perp$ falls off rapidly in the regime of relativistic motion.

The behavior of the longitudinal mode is starkly different. We can obtain a first matching condition by integrating \cref{eq:condition} across the wall:
\begin{equation}
	\lim_{\epsilon \rightarrow 0} \int_{- \epsilon}^{+ \epsilon} dz \, \partial_\mu \left( m_V^2 (z) V^\mu_l \right) = 0
	\qquad \Rightarrow \qquad
	m_V^2(z) V^3_l \ \text{is continuous at} \ z = 0 \ .
\label{eq:cont_1}
\end{equation}
Integrating \cref{eq:EOM} provides the second condition we need. Since $\partial_\mu V_l^\mu$ no longer vanishes, expanding Eq.(\ref{eq:EOM}) we now have $(\Box + m_V^2(z)) V_l^\mu - \partial^\mu (\partial_\nu V_l^\nu) = 0$. The matching condition arising from the $\mu = 0$ equation is degenerate with \cref{eq:cont_1}. % for the step function profile.
Instead, focusing on $\mu=3$, we find
\begin{equation}
	\lim_{\epsilon\rightarrow 0} \int_{- \epsilon}^{+\epsilon} dz \, \left( (\Box + m_V^2(z)) V_l^3 + \partial_z (\partial_\nu V_l^\nu) \right) = 0 \qquad \Rightarrow \qquad
	\partial_t V_l^0 \ \ \text{is continuous at } z= 0 \ .
\label{eq:A^0_z0}
\end{equation}
Given that the time dependence of $V^\mu$ is of the form $e^{-i \omega t}$ for all $z$, the previous requirement is equivalent to demanding that $V_l^0$ itself remains continuous.
The corresponding reflection probability now reads
\begin{equation}
	R_l = |r_l|^2 = \left| \frac{\tilde{m}^2 \, k_z - m^2 \, \tilde{k}_z}{\tilde{m}^2 \, k_z + m^2 \, \tilde{k}_z} \right|^2 \xrightarrow{\omega \gg m, \tilde{m}} \left( \frac{ \tilde{m}^2 - m^2 }{ \tilde{m}^2 + m^2 } \right)^2       
    \simeq \left( \frac{ \Delta m^2 }{ 2 m^2 } \right)^2 \ ,
\label{eq:RL}
\end{equation}
where the last step assumes $\Delta m^2 \ll m^2 \simeq \tilde m^2$.
As advertised in the Introduction, in the regime of relativistic motion the longitudinal reflection probability approaches a constant, independent of $\gamma$.
\footnote{At the level of a St\"uckelberg theory, a discontinuity in the number of degrees of freedom (dof) occurs in the limit $m \rightarrow 0$, with 2 vs 3 physical polarizations at either side of the interface. The fact that $R_l \rightarrow 1$ as $m \rightarrow 0$ in \cref{eq:RL} reflects the observation that the longitudinal mode would be unphysical in the region $z<0$, and therefore must not propagate into the region $z >0$.
In an Abelian Higgs UV-completion, the number of dof of course stays continuous, with the would-be longitudinal accounted for by the appropriate linear combination of the two real dof of the complex Higgs.
The physical dof are two transverse modes and the suitable combination of the real scalars, and therefore the treatment presented in this section is not applicable in the massless regime.
Scattering with $m=0$ has been studied long ago, e.g.~\cite{Farrar:1994vp}.
As emphasized in the Introduction, and in \cref{sec:model}, we instead focus on cases where (in the Abelian Higgs language) the Higgs vev is `on' on both sides of the wall.}
\footnote{See \cref{sec:get} for an alternative derivation of \cref{eq:RL} making use the of Goldstone nature of longitudinals at high energies.}

%%%%%%%%%%%%%%%%%%%%%%%%%%%%%%%%%%%%%%%%%
\subsection{\label{sec:smooth} \ A smooth wall}

In a realistic situation where the wall thickness is finite and the vector mass varies smoothly, the analytic results of the previous subsection only provide an accurate approximation to the reflection and transmission probabilities in the regime $\omega \ll L^{-1}$. In what follows, we perform a more general analysis of the case of finite width. We parametrize the dark photon mass as in \cref{eq:mass_step}, except now
\begin{equation}
    \Delta m_V^2 (z) = \Delta m^2 \Theta_L (z) \ ,
\end{equation}
where $\Theta_L (z)$ is no longer a step function, but rather some smooth function that approaches 0 and 1 for large negative and positive $z$ respectively, and with an appropriate step-function limit as $L \rightarrow 0$. We will discuss a specific choice of mass profile at the end of this subsection, but will otherwise keep things general.

Let us start by building some intuition behind the radically different behavior exhibited by the transverse and longitudinal components discussed in \ref{sec:step} by taking a closer look at the corresponding field equations. It will be helpful to factor out the time dependence of $V^\mu$, as follows
\begin{equation}
	V^\mu (t,z) = v^\mu (z) e^{- i \omega t} \  ,
\end{equation}
and instead focus on the behavior of the $v^\mu(z)$. As we discussed previously, the equations of motion for the transverse modes read $(\Box + m_V^2(z)) V_\perp^\mu = 0$. In terms of the $v^\mu$, this can be written as
\begin{align}
	\left( \partial_z^2 + k_z^2 \right) v^\mu_\perp (z) = \Delta m^2_V (z) \, v^\mu_\perp (z) \ ,
\label{eq:full_perp}
\end{align}
with $k_z^2 = \omega^2 - m^2$.
The transverse components obey a Schr\"odinger-like equation for a particle moving in one dimension with ``energy" $k_z^2$ in the presence of a potential $U_\perp (z) = \Delta m_V^2 (z)$. In the $L \rightarrow 0$ limit, our problem reduces to one-dimensional quantum mechanical scattering on a step potential, with energies above the step. Indeed, \cref{eq:Rperp} is just the reflection probability for this classic problem.

On the other hand, the Schr\"odinger equation governing the behavior of the longitudinal component reads
\begin{align}
	\left( \partial_z^2 + k_z^2 \right) \lambda (z) = U_l(z) \lambda (z) 
    \qquad \text{with} \qquad
    U_l(z) = \Delta m_V^2 (z) + \frac{3}{4} \left( \frac{\partial_z m_V^2 (z)}{m_V^2 (z)} \right)^2 - \frac{1}{2} \frac{\partial^2_z m_V^2(z)}{m_V^2(z)}  \ ,
\label{eq:full_long}
\end{align}
where we have defined
\begin{equation}
    \lambda (z) \equiv \frac{m_V(z)}{\omega} v^3_l(z) \qquad \text{such that} \qquad
    \lambda (z) \rightarrow \begin{cases} 	e^{i k_z z} + r_l e^{- i k_z z}  & \quad \text{for} \qquad z \ll - L \\
								t_l e^{i \tilde{k}_z z} & \quad \text{for} \qquad z \gg + L \end{cases} \ .
\end{equation}
The effective scattering potential for the longitudinal component, $U_l (z)$, depends on the length scale over which the mass changes not just through the choice of mass profile but through its derivatives -- much unlike its transverse counterparts. It is this crucial difference that leads to the strikingly different behavior of longitudinal and transverse modes.

In the relativistic limit, $k_z \simeq \omega \gg m$, we might neglect the first term in $U_l(z)$. The derivative terms in the effective potential are localized in a region of thickness $\sim L$, and the second-derivative term is dominant whenever $\Delta m^2 / m^2$ is tiny. We then have
\begin{equation}
    U_l(z)  \simeq - \frac{1}{2} \left( \frac{\partial^2_z m_V^2 (z)}{m_V^2 (z)} \right)
            \simeq - \frac{\Delta m^2}{2 m^2} \Theta_L'' (z) \ .
\label{eq:Vl_smooth}
\end{equation}
The step-function limit discussed in the previous subsection corresponds to $\Theta_L'' (z) \xrightarrow{L \rightarrow 0} \delta' (z)$, and \cref{eq:full_long} reduces to the Schr\"odinger equation in a potential $U_l(z) = - \frac{\Delta m^2}{2 m^2} \delta' (z)$. Solving this equation subject to the appropriate boundary conditions, one indeed recovers the result of \cref{eq:RL} to leading order in the ratio $\Delta m^2 / m^2$, as we summarize in appendix~\ref{app:deltaprime}.

Since the longitudinal reflection coefficient in the relativistic regime is $R_l \ll 1$ whenever the change in the dark photon mass is tiny, we can go further and leverage the one-dimensional Born approximation familiar from quantum mechanics to obtain an analytic expression for the reflection probability in the case of a smoothly varying mass.
As we summarize in appendix~\ref{app:born}, the Born approximation to the longitudinal reflection coefficient can be written in closed form in terms of the effective scattering potential in \cref{eq:full_long} as
\begin{equation}
    R_{l\text{,\,Born}} = \frac{1}{4k_z^2} \left| \int_{-\infty}^{\infty} dz \ e^{2ik_z z} \ U_l(z) \right|^2 \ .
    \label{eq:born_prob}
\end{equation}
Plugging \cref{eq:Vl_smooth} into \cref{eq:born_prob} and integrating by parts, we find
\begin{align}
    R_{l\text{,\,Born}} 
    &\simeq \frac{1}{4k_z^2} \left( \frac{\Delta m^2}{2 m^2} \right)^2
        \left| 
            \Big[ e^{2ik_z z} \ \Theta_L' (z) \Big]_{-\infty}^{+\infty} - 2ik_z \int_{-\infty}^{\infty} dz \ e^{2ik_z z} \ \Theta_L' (z) 
        \right|^2 \\
    &\simeq \left( \frac{\Delta m^2}{2 m^2} \right)^2
        \left| 
            \int_{-L}^{L} dz \ \left( 1 + 2ik_z z + \mathcal{O} (k_z^2 z^2) \right) \ \Theta_L' (z) 
        \right|^2 \\
    &= \left( \frac{\Delta m^2}{2 m^2} \right)^2 \left( 1 + \mathcal{O}(k_z^2 L^2) \right) \ . \label{eq:RlBorn}
\end{align}
To get the second line, we used that $\Theta_L'$ quickly vanishes for $|z| \gtrsim L$, and we Taylor expanded the exponential for $|k_z z| \ll 1$. Since $\Theta_L$ interpolates between 0 and 1, $\Theta_L '$ can be treated as a probability density function. With this interpretation, the integral in the second line is essentially an average of $1 + 2i k_z z$ over the region $(-L,L)$, weighted by this probability density. The average of $k_z z$ is bounded in magnitude by $k_z L$, giving the final line.
Eq.~(\ref{eq:RlBorn}) coincides with the step-function result of \cref{eq:RL} provided $k_z \simeq \omega \ll L^{-1}$, as expected. Moreover, it highlights how the existence of a kinematic regime where the longitudinal reflection coefficient stays nearly constant is in fact a generic feature of any smooth mass profile, lasting all the way up to $k_z \simeq \omega \sim L^{-1}$.

Given a specific profile, we can use \cref{eq:born_prob} to obtain an analytic approximation to the reflection probability. For example, for a wall profile of the familiar kink form
\begin{equation}
\label{eq:thetaL_specific}
    \Theta_L (z) = \frac{1}{4} \left[1+\tanh(z/L) \right]^2 \ ,
\end{equation}
one finds
\begin{align}
\label{eq:Born_specific}
    R_{l\text{,\,Born}} \simeq
    \left( \frac{\Delta m^2}{2 m^2} \right)^2
    \frac{\pi^2 (k_z L)^2 \left[ 1 + (k_z L)^2 \right]}{\sinh^2 (\pi k_z L)} \ .
\end{align}
Eq.~(\ref{eq:Born_specific}) reproduces the result of \cref{eq:RL} in the regime $\omega \simeq k_z \ll L^{-1}$, up to corrections of $\mathcal{O} (k_z^2 L^2)$, whereas in the ultra-relativistic limit it takes the form
\begin{equation}
    R_{l\text{,\,Born}} \simeq \left( \frac{\Delta m^2}{2 m^2} \right)^2 \times  4 \pi^2 \left( k_z L \right)^4 e^{- 2 \pi k_z L} \qquad \text{for} \qquad k_z \simeq \omega \gg L^{-1} \ .
\end{equation}
Indeed, the longitudinal reflection probability dies off exponentially fast -- as expected -- in the regime of ultra-relativistic motion.

\begin{figure}[h]
  \centering
  \includegraphics[width=\textwidth]{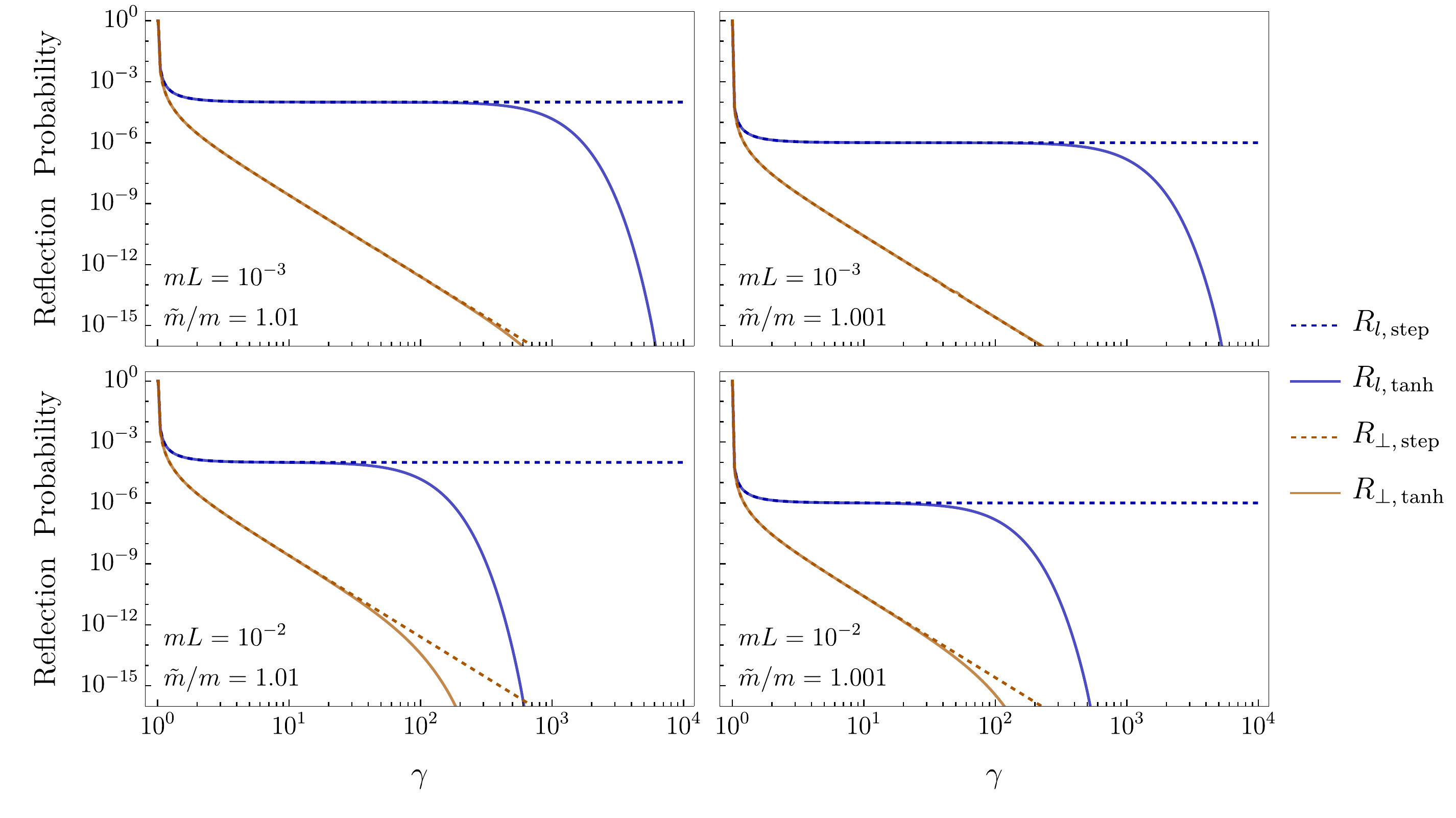}
  \caption{Reflection probability for longitudinal (blue) and transverse (orange) massive vectors scattering off a planar wall in the limit of normal incidence, as a function of the Lorentz $\gamma$-factor. The change in mass across the interface is $1\%$ and $0.1\%$ for the left and right panels respectively. Solid lines show numerical results for a smooth mass profile of the form $m_V^2 (z) = m^2 + \Delta m^2 \Theta_L(z)$, with $\Theta_L(z)$ as in \cref{eq:thetaL_specific}. Dashed lines correspond to the step function results of \cref{eq:Rperp} and (\ref{eq:RL}). As anticipated, the step-function analysis provides an excellent approximation up to $\gamma \sim (m L)^{-1}$. Details on how we obtain our numerical results are summarized in appendix~\ref{app:numerics}.
  }
  \label{fig:coeffs}
\end{figure}

The results of this subsection are best summarized in \cref{fig:coeffs}, where we show the reflection probability for longitudinal and transverse vectors for the mass profile in \cref{eq:thetaL_specific}, obtained by numerically solving the corresponding equations of motion.
As advertised, the longitudinal reflection coefficient features a plateau for $\gamma$-factors in the regime $1 \ll \gamma \ll (L m)^{-1}$, which is well-approximated by the step-function result of \cref{eq:RL}.
The consequences of this behavior for the evolution of bubble walls are the topic of the next section.
%%%%%%%%%%%
%%%%%%%%%%%

%%%%%%%%%%%%%%%%%%%%%%%%%%%%%%%%%%%%%%%%%
\section{\label{sec:friction} \ Dark photon friction on bubble walls}
%%%%%%%%%%%%%%%%%%%%%%%%%%%%%%%%%%%%%%%%%

We now discuss the implications of the results presented in \cref{sec:probs} for the dynamics of expanding vacuum bubbles in the early Universe.
In~\cref{sec:mdp}, we compute the pressure on an expanding wall due to the presence of dark photons, and argue that the requirement for run-away walls can be much stronger than previously believed.
Section~\ref{sec:gamma_eq} focuses on the dynamics of bubble walls that reach an equilibrium regime as a result of longitudinal friction, including a self-consistent determination of the equilibrium $\gamma$-factor.
Once in equilibrium, most of the energy released in the transition goes into making a fraction of the dark photons relativistic.
The fate of this dark radiation depends sensitively on the size of self-interactions among the dark photons, as well as between the dark photons and particles in the thermal plasma, as we discuss in~\cref{sec:fate_dr}.
If interaction rates are negligible -- an assumption more likely to hold if the sector undergoing vacuum decay is cold -- we argue that the reflected dark photons accumulate in a thin ``shell" of dark radiation surrounding the expanding bubbles. If they remain relativistic until late times, their contribution to $\Delta N_\text{eff}$ could be observable for phase transitions with strength $\alpha \sim 10^{-2} - 10^{-1}$.
Alternatively, if the scalar sector is in equilibrium with the thermal plasma, interactions between the reflected dark photons and $\phi$ particles can easily be efficient, in which case the energy density in dark radiation gets transferred instead into the thermal fluid.

%%%%%%%%%%%%%%%%%%%%%%%%%%%%%%%%%%%%%%%%%
\subsection{ \ Maximum Dynamic Pressure}
\label{sec:mdp}

It is helpful to rearrange \cref{eq:Pgamma} using the relationship $T_\lambda = 1 - R_\lambda$, as follows:
\begin{equation}
	\mathcal{P} = \gamma | \vec v | n_V \left\{ \frac{1}{3} R_l (k_z + \tilde{k}_z) + (k_z - \tilde{k}_z) + \frac{2}{3} R_\perp (k_z + \tilde{k}_z) \right\} \ .
\label{eq:pressure_v2}
\end{equation}
In the relativistic limit, $k_z + \tilde{k}_z \simeq 2 \gamma m$ and $k_z - \tilde{k}_z \simeq \Delta m^2/(2 \gamma m)$, and \cref{eq:pressure_v2} reads
\begin{equation}
	\mathcal{P} \simeq \frac{2}{3} \gamma^2 \rho_V R_l + \rho_V \frac{\Delta m^2}{2m^2} + \frac{4}{3} \gamma^2 \rho_V R_\perp
	\qquad \text{for} \qquad \gamma \gg 1 \ .
\label{eq:Prel}
\end{equation}
The second term above is just the asymptotic contribution of \cref{eq:P_UR}. 
The last term corresponds to reflections of transverse modes. As discussed around \cref{eq:Rperp}, $R_\perp \propto \gamma^{-4}$ already in the inter-relativistic regime, and therefore this third term falls off as $\gamma^{-2}$ for large $\gamma$. In contrast, the peculiar behavior of the longitudinal reflection probability, $R_l$, which stays approximately constant in the region of inter-relativistic motion, leads to a contribution to the overall pressure on the expanding interface that grows $\propto \gamma^2$. In total, neglecting the last term above and substituting the expression for $R_l$ appropriate in the inter-relativistic regime (remember~\cref{eq:RL}), we have:
\begin{align}
	\mathcal{P}
        \simeq
        \underbrace{ \frac{2}{3} \gamma^2 \rho_V \left( \frac{\Delta m^2}{2 m^2} \right)^2 }_\text{longitudinal reflections} + \
        \underbrace{\rho_V \frac{\Delta m^2}{2m^2}}_{\mathcal{P}_\infty}
	\qquad \text{for} \qquad1 \ll \gamma \ll (m L)^{-1} \ .
\label{eq:Pinter}
\end{align}
We emphasize that this $\gamma^2$-growing pressure is a transient phenomenon that is only present for $\gamma$-factors within the range indicated in \cref{eq:Pinter}. Once $\gamma \gg (m L)^{-1}$, the longitudinal reflection probability dies off exponentially fast, leaving the second term above as the sole significant contribution to the overall pressure.

A consequence of our previous discussion is that the pressure exerted on the expanding interface will reach a maximum near the end of the inter-relativistic regime, before dropping down to $\mathcal{P}_\infty \simeq \rho_V \times \Delta m^2 / 2m^2$ in the ultra-relativistic limit. As advertised in the Introduction, we will refer to this pressure peak as Maximum Dynamic Pressure.
Parametrically, it is given by
\begin{align}
	\mathcal{P}_\text{mdp}
        \simeq \frac{2}{3} \gamma_\text{max}^2 \rho_V R_l + \rho_V \frac{\Delta m^2}{2m^2}
		\sim \frac{\rho_V}{(L m)^2} \left( \frac{\Delta m^2}{2 m^2} \right)^2 + \rho_V \frac{\Delta m^2}{2 m^2} \ ,
        \label{eq:mdp}
\end{align}
where $\gamma_\text{max} \sim (mL)^{-1}$ is the value of $\gamma$ at Maximum Dynamic Pressure. As a result, for bubble walls to become run-away, the following condition must be satisfied:
\begin{equation}
	\text{Run-away criterion:}
    \qquad \qquad \Delta V > \mathcal{P}_\text{mdp}
    \qquad \qquad \text{(this work)} \ .
\label{eq:maxq}
\end{equation}
Eq.(\ref{eq:maxq}) replaces \cref{eq:runaway_naive} as a diagnostic of run-away bubble walls in the presence of phase-dependent massive dark photons in transitions with access to the regime of inter-relativistic motion, and is a primary new result of this work.

If the first term in \cref{eq:mdp} is a small correction on top of the asymptotic pressure, $\mathcal{P}_\infty$, the effect of longitudinal mode reflections will be largely irrelevant to describe the dynamics of the expanding bubbles. However, if the first term dominates then \cref{eq:maxq} can be much stronger than the requirement $\Delta V > \mathcal{P}_\infty$ quoted in \cref{eq:runaway_naive}.
Parametrically, the effect of longitudinal reflections will dominate the overall pressure on the expanding interface provided
\begin{equation}
\label{eq:mdp_long_condition}
	 m L  \ll \sqrt{\frac{\Delta m^2}{m^2}} \ .
\end{equation}
When $\Delta m^2 /m^2 \ll 1$, this is a stronger requirement than the condition in \cref{eq:def_ir} for the inter-relativistic regime to be accessible, although the main feature in both cases is that there must be a significant hierarchy between the mass of the dark photon and the energy scale $L^{-1}$ characterizing the thickness of the bubble wall.

Our discussion thus far has only been concerned with the pressure due to a population of cold dark photons. 
If the sector undergoing the transition is cold, such that vacuum decay proceeds via quantum tunneling and there is no thermal plasma that interacts with the bubble walls, this will be a good approximation to the overall pressure.
Alternatively, if the scalar sector is in equilibrium with the thermal fluid, an additional source of pressure will be present due to interactions between the plasma and the wall.
As mentioned in the Introduction, particles in the plasma that gain mass across the wall contribute as $\mathcal{P}_\infty \sim \Delta m^2 T^2$ in the relativistic limit~\cite{Bodeker:2009qy}.
In perturbative theories this source of friction can easily fall below $\Delta V$, and therefore won't be large enough to obstruct the acceleration of the bubble walls.
Potentially more significant are those cases where the phase transition sector features gauge bosons that acquire a mass as they cross the wall.
As anticipated in \cref{sec:intro}, this can lead to an additional source of pressure from the transition radiation emitted as charged particles cross the interface, of the form $\mathcal{P}_\infty \sim \gamma g^2 \Delta m_v T^3$~\cite{Bodeker:2017cim,Azatov:2020ufh,Gouttenoire:2021kjv}.
For this source of friction to be subdominant to that from longitudinal dark photons, one would need
\begin{equation}
	\gamma_\text{eq} g^2 \Delta m_v T_*^3 \lesssim \Delta V \ .
\end{equation}
This condition can be interpreted as an upper bound on $T_*$ relative to the typical energy scale of the phase transition.
Parametrically, taking for simplicity $\Delta V \sim v^4$ and $\Delta m_v \sim g v$, we find
\begin{align}
	\frac{T_*}{v} 	& \lesssim  \frac{1}{g \, \gamma_\text{eq}^{1/3}} \\
				& \sim 10^{-3} \, \left( \frac{1}{g} \right)
		\left( \frac{\Delta m^2 / m^2}{10^{-4}} \right)^{1/3} \left( \frac{10^{-2}}{\alpha} \right)^{1/6} \left( \frac{100 \ \text{GeV}}{T_*} \right)^{1/6} \left( \frac{\rho_\text{dm}}{\rho_V} \right)^{1/6} \ .
\end{align}
If the relevant gauge couplings are $g = \mathcal{O}(1)$ -- as in the Standard Model -- then neglecting transition radiation would require $T_* \ll v$, i.e.~the transition needs to be significantly super-cooled. More generally, in hidden sectors where the relevant gauge couplings are $g \ll 1$, the above condition could be satisfied even within `standard' thermal transitions where $T_* \sim v$.
A more comprehensive analysis of the class of thermal transitions for which this assumption holds is an interesting direction for future investigation.

\begin{figure}[h]
  \centering
  \includegraphics[width=\textwidth]{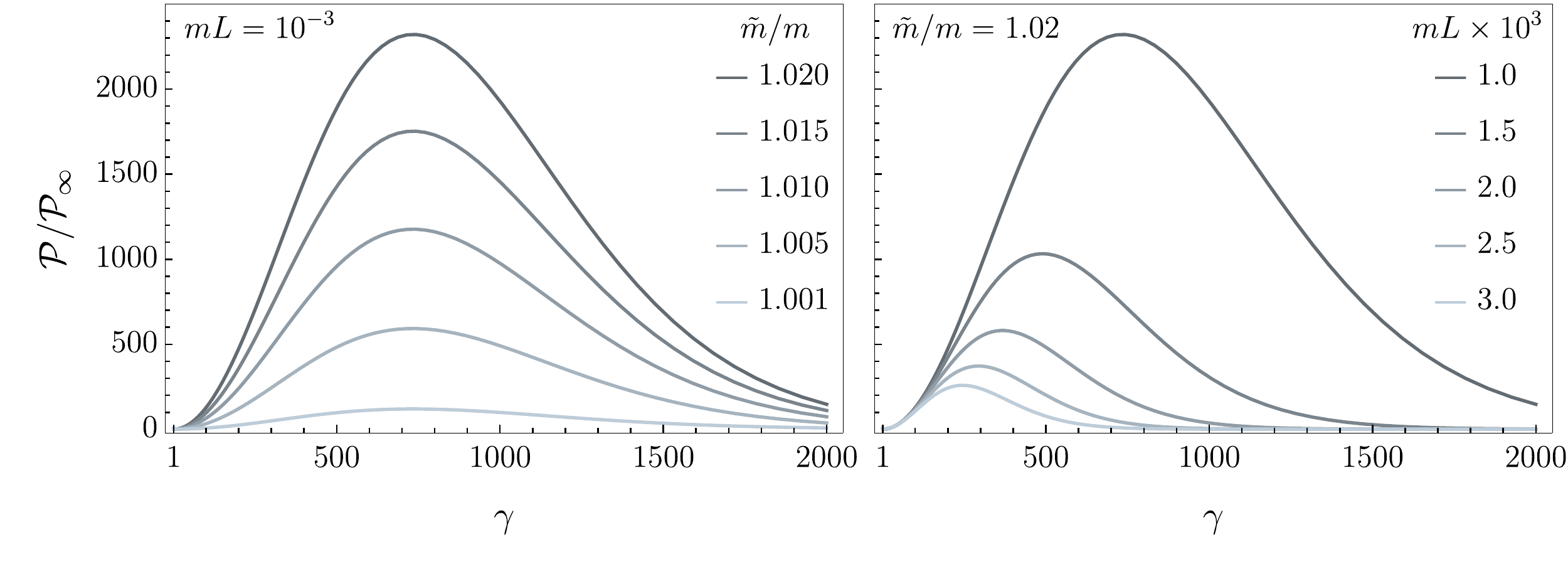}
  \caption{ The pressure exerted by massive dark photons on an expanding bubble wall can feature a non-monotonic dependence on the wall $\gamma$-factor due to reflections of longitudinal modes, reaching a peak at $\gamma \sim (mL)^{-1}$ that is potentially much larger than its asymptotic value in the ultra-relativistic limit. The criterion for run-away bubble walls can therefore be far stronger than simply requiring $\Delta V > \mathcal{P}_\infty$, as discussed around \cref{eq:maxq}. {\bf Left:} increasing the relative change in the dark photon mass raises the Maximum Dynamic Pressure without significantly altering its position. {\bf Right:} decreasing the wall thickness increases \emph{both} the height and position of the MDP. This behavior is in qualitative agreement with the discussion in and around \cref{eq:mdp}. Both plots have been obtained by evaluating \cref{eq:pressure_v2} after numerically obtaining reflection coefficients corresponding to the smooth wall profile of \cref{eq:thetaL_specific}.}
  \label{fig:pressure}
\end{figure}

Fig.~\ref{fig:pressure} shows the pressure on an expanding interface due to massive dark photons, relative to its asymptotic value in the limit $\gamma \rightarrow \infty$, in cases where the inter-relativistic kinematic regime identified in \cref{eq:def_ir} is accessible during the wall's expansion.
The evolution of bubble walls that fail the run-away condition of \cref{eq:maxq} as a result of longitudinal friction is the topic to which we now turn.
%%%%%%%%%%%
%%%%%%%%%%%

%%%%%%%%%%%%%%%%%%%%%%%%%%%%%%%%%%%%%%%%%
\subsection{ \ Equilibrium $\gamma$-factor and energy budget}
\label{sec:gamma_eq}

If the Maximum Dynamic Pressure is dominated by the reflection of longitudinal modes, and the run-away criterion of \cref{eq:maxq} is not satisfied, then bubble walls will reach an equilibrium regime once $\mathcal{P}(\gamma_\text{eq}) \simeq \Delta V$.
Parametrically, the equilibrium $\gamma$-factor is given by
\begin{align}
     	\gamma_\text{eq} 	& \simeq \left( \frac{3 \Delta V}{2 \rho_V R_l} \right)^{1/2}	\label{eq:gammastar} \\
					& \sim  10^9 \left( \frac{10^{-4}}{\Delta m^2 / m^2} \right) \left( \frac{\alpha}{10^{-2}} \right)^{1/2} \left( \frac{T_*}{100 \ \text{GeV}} \right)^{1/2} \left( \frac{\rho_\text{dm}}{\rho_V} \right)^{1/2} \ . \label{eq:gammastar_num}
\end{align}

How easy is it for bubble walls to reach an equilibrium regime as a result of longitudinal reflections?
An obvious self-consistency condition on our determination of $\gamma_\text{eq}$ in \cref{eq:gammastar} is that it lies below the $\gamma$-factor at Maximum Dynamic Pressure,~i.e.~
\begin{equation}
    \gamma_\text{eq} \lesssim \frac{1}{m L} \ .
    \label{eq:gammaeq_belowmax}
\end{equation}
Additionally, a ``kinematic" condition is that the expanding walls reach equilibrium \emph{before} the bubble walls collide and the phase transition ends. The bubble radius at collision is set by the Hubble scale at the epoch of the phase transition. Demanding that the size of the expanding bubbles at the onset of the equilibrium regime is a fraction $x<1$ of the Hubble radius, we find
\begin{equation}
    R_\text{eq} \simeq \gamma_\text{eq} R_0 \lesssim x H(T_*)^{-1}
    \qquad \Rightarrow \qquad
    \gamma_\text{eq} \lesssim \frac{x}{R_0 H(T_*)} \ ,
    \label{eq:gammaeq_belowcoll}
\end{equation}
where $R_0$ is the critical bubble radius.

\begin{figure}[h]
  \centering
  \includegraphics[scale=0.9]{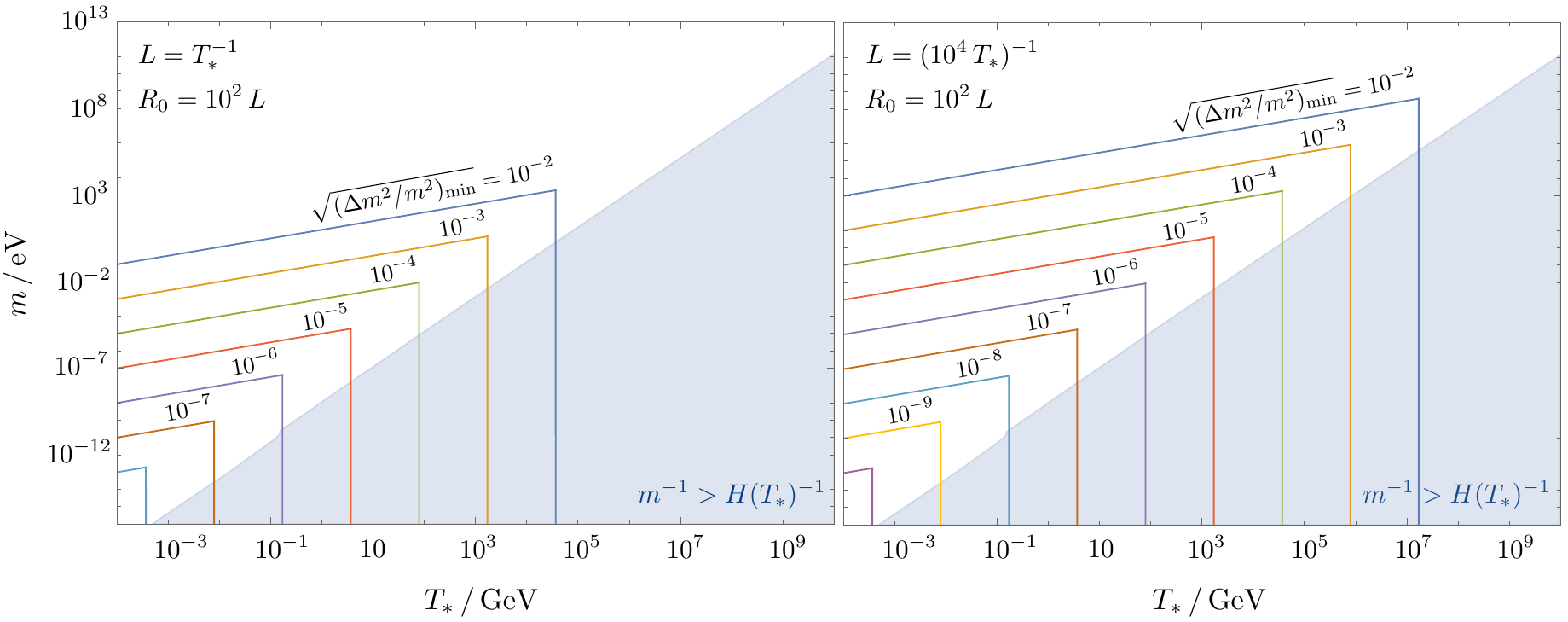}
  \caption{Contour lines indicating the lower bound on the fractional change of the dark photon mass for longitudinal reflections to create enough friction to halt the acceleration of the bubble walls. These contours saturate the inequality in \cref{eq:deltamin}, with the choice $x = 10^{-1}$ for illustration. The energy density in cold dark photons is taken to be that of the dark matter, and the difference in energy densities, $\Delta V$, is taken to be $1 \%$ of that in the Standard Model bath (i.e.~$\alpha = 10^{-2}$ in \cref{eq:alpha_def}). The blue shaded area corresponds to dark photon masses such that the corresponding Compton wavelength exceeds the Hubble radius, $m^{-1} > H(T_*)^{-1}$ -- a regime that lies outside the validity of our discussion. {\bf Left:} The wall thickness is related to the temperature of the thermal plasma at the phase transition epoch as $L=T_*^{-1}$, as expected in a `standard' thermal transition. {\bf Right:} $L=(10^4 \, T_*)^{-1} \ll T_*^{-1}$, as could be the case if the phase transition took place within a cold hidden sector with a characteristic energy scale $\gg T_*$, or for a thermal transition featuring significant super-cooling (recall discussion below \cref{eq:L_dim4}).
  }
  \label{fig:models}
\end{figure}

Eq.(\ref{eq:gammaeq_belowmax}) and (\ref{eq:gammaeq_belowcoll}) can be interpreted in various ways, but perhaps the most relevant for us is to regard them as \emph{lower bounds} on the fractional change of the dark photon mass for the effect of longitudinal pressure to be large enough to stop the acceleration of the bubble walls, as follows:
\begin{equation}
	\frac{\Delta m^2}{m^2} \gtrsim \left( \frac{\Delta m^2}{m^2} \right)_\text{min}
	\equiv 2 \left( \frac{3 \Delta V}{2 \rho_V} \right)^{1/2} \, \times  \, \text{Max} \left\{ m L  , \, x^{-1} R_0  H(T_*) \right\} \ .
	\label{eq:deltamin}
\end{equation}
This is illustrated in \cref{fig:models}, where we show contours of $\sqrt{(\Delta m^2 / m^2)_\text{min}}$ for various choices of the underlying parameters, as described in the caption.
As can be seen in \cref{fig:models}, even extremely small changes in the mass of the dark photon across the interface can cause enough friction to halt the acceleration of the bubble walls.
%%%%%%%%%%%%%%%%%%%%%%%%%%%%%%%%%%%%%%%%%
%%%%%%%%%%%%%%%%%%%%%%%%%%%%%%%%%%%%%%%%%

After the bubble walls reach an equilibrium speed, they carry a decreasing fraction of the total energy available in the transition:
\begin{equation}
	\frac{E_\text{wall}}{E_\text{total}} \simeq \frac{4 \pi R(t)^2 \gamma_\text{eq} \sigma}{\frac{4 \pi}{3} R(t)^3 \Delta V} \sim \frac{\gamma_\text{eq} \sigma}{\Delta V \, R(t)} \ ,
\end{equation}
where $R(t)$ is the bubble radius at time $t$. As $R(t)$ grows, $E_\text{wall} / E_\text{total}$ quickly becomes tiny. By comparison, the above ratio is identically 1 for transitions in vacuum, where all the available energy goes into accelerating the bubble walls~\cite{Coleman:1977py}. In thermal transitions where the walls reach an equilibrium speed due to friction from the thermal plasma, most of the available energy goes instead into producing motion in the form of sound waves or hydrodynamic turbulence, as mentioned in the Introduction and summarized in~\cite{Caprini:2015zlo,Caprini:2019egz}.
In the case at hand, as we will now show, the energy released in the transition goes instead into accelerating a fraction of the dark photons, turning them into dark radiation.

In the rest frame of the bubble wall, reflected longitudinal modes propagate away from the interface with Lorentz-factor $\gamma_\text{eq}$, while in the rest frame of the dark photons far away from the wall -- loosely, the rest frame of the dark matter -- the $\gamma$-factor of the reflected dark photons is given by
\begin{equation}
    \gamma_\text{dr} = \frac{1 + | \vec v_\text{eq} |^2}{1 - | \vec v_\text{eq} |^2} \simeq 2 \gamma_\text{eq}^2 \gg \gamma_\text{eq} \ .
    \label{eq:vdr}
\end{equation}
Within a Hubble volume, the average number density of dark photons that become relativistic as the bubble walls sweep across the dark matter from the onset of the equilibrium regime until the bubbles collide can be approximated by
\begin{equation}
	\langle n_\text{dr} \rangle \simeq \frac{1}{3} R_l n_V \left( 1 - \frac{R_\text{eq}^3}{R_\text{coll}^3} \right) \ .
\label{eq:ndr}
\end{equation}
The factor in parenthesis takes care of the fact that we are only interested in keeping track of the fraction of dark matter that gets converted into dark radiation during equilibrium (before, most of the energy released as the bubbles grow goes into accelerating the bubble walls).
Obviously, if the bubble walls only reach equilibrium right before they collide, $R_\text{eq} \approx R_\text{coll}$, then $\langle n_\text{dr} \rangle$ will be correspondingly tiny.
However, provided the onset of equilibrium takes place well before collision, then $\langle n_\text{dr} \rangle \simeq \frac{1}{3} R_l n_V$.
For example, taking $R_\text{eq} \simeq \gamma_\text{eq} R_0$ and $R_\text{eq} \simeq x H(T_*)^{-1}$, we find $R_\text{eq}^3 / R_\text{coll}^3 \sim 10^{-15} \ll 1$, where we have evaluated all parameters as in \cref{eq:gammastar_num}, and for illustration we have taken $x = 10^{-2}$ as well as $R_0^{-1} = T_* = 100 \ \text{GeV}$.
More generally, notice that having $R_\text{eq} \lesssim R_\text{coll}$ is a requirement for dark photon reflections to affect significantly the evolution of the bubble walls (as per our discussion around \cref{eq:gammastar_num}), and so we proceed under this assumption.

Right after bubble walls collide, the volume-averaged energy density in dark radiation is therefore
\begin{equation}
	\rho_\text{dr} = \omega_\text{dr} \, \langle n_\text{dr} \rangle \simeq \frac{2}{3} \gamma_\text{eq}^2 \, R_l \, \rho_V \qquad \text{at} \qquad T \sim T_* \ ,
\label{eq:rhodr}
\end{equation}
where $\omega_\text{dr} = \gamma_\text{dr} m$ and $\gamma_\text{dr}$ is given in \cref{eq:vdr}.
Notice the right-hand-side above is equal to $\Delta V$ by \cref{eq:gammastar} -- consistent with energy conservation in the rest frame of the dark matter.
Thus, in phase transitions where bubble walls reach an equilibrium regime as a result of friction from the dark photons, most of the difference in vacuum energy densities goes into turning a fraction of the (cold) dark photons into dark radiation.

%%%%%%%%%%%%%%%%%%%%%%%%%%%%%%%%%%%%%%%%%
\subsection{ \ Fate of the dark radiation}
\label{sec:fate_dr}

The fate of the reflected dark photons depends sensitively on the values of the various underlying parameters. As a result, general statements about the later evolution of the dark radiation are not possible, and a comprehensive study spanning all available parameter space is well beyond the scope of this work. Instead, we will focus on highlighting the possible outcomes given certain assumptions.

Although sweeping statements are not possible, there are two qualitatively different cases that merit separate consideration, depending on whether the sector undergoing vacuum decay is part of the thermal plasma or belongs in a cold hidden sector such that the transition proceeds via quantum tunneling. We will focus on the cold scenario first.
In this case, the absence of a thermal population of particles that interact with the wall background leaves the dark photons are the sole source of pressure on the expanding interface, and the evolution of the dark radiation as the bubbles grow depends primarily on the strength of the self-interactions among the dark photons.
As discussed in \cref{sec:model}, quartic dark photon self-interactions are generally part of the low energy theory, and in the context of an Abelian Higgs UV-completion the relevant quartic coupling is as given in \cref{eq:lambdav}. Interactions among the dark radiation and the cold dark photons occur at a center-of-mass energy $\sqrt{s} / 2 \simeq \gamma_\text{eq} m$, which always fall below the cutoff scale of the effective theory. In this kinematic regime, the relevant cross section is as given in \cref{eq:sigma_smalls}, and one finds that the corresponding interaction rate can be tiny compared to $H(T_*)$ across significant portions of parameter space.

When interactions between the dark matter and the reflected dark photons are negligible, the following picture emerges.
Going back to \cref{eq:vdr}, notice that (in the dark matter frame) the speed of the dark radiation is ever-so-slightly larger than the speed of the bubble wall:
\begin{align}
	\Delta v_\text{dr} & \equiv |\vec v_\text{dr}| - |\vec v_\text{eq}| = |\vec v_\text{eq}| \frac{1 - |\vec v_\text{eq}|^2}{1 + |\vec v_\text{eq}|^2} \simeq \frac{|\vec v_\text{eq}|}{2 \gamma_\text{eq}^2} \ll 1 \ .
\end{align}
As a result, from the beginning of the equilibrium regime until the moment of bubble wall collisions, reflected dark photons become uniformly distributed on a shell of thickness $\Delta L$ in front of the interface, with
\begin{equation}
	\Delta L \sim \Delta v_\text{dr} \times \Delta t \simeq \frac{|\vec v_\text{eq}| \Delta t}{2 \gamma_\text{eq}^2} \lesssim \frac{R_\text{coll}}{2 \gamma_\text{eq}^2} \ll R_\text{coll} \ ,
    \label{eq:DR_layer}
\end{equation}
where $\Delta t$ refers to the time interval between when equilibrium is reached and collision of the bubble walls, and we have used $ |\vec v_\text{eq}| \Delta t \lesssim R_\text{coll}$.
Thus, as the bubbles grow bigger, a shell of dark radiation forms, moving ever-so-slightly in front of the bubble walls. Although the walls move at constant speed, most of the energy density remains localized in a thin layer close to the surface of the expanding bubbles.

The fate of these shells of radiation depends on the interactions between relativistic dark photons as the bubble walls meet. Now, the relevant center-of-mass-energy is $\sqrt{s} / 2 \sim \omega_\text{dr} \simeq 2 m \gamma_\text{eq}^2$, and interactions are often well-described within the high energy Goldstone regime of \cref{eq:sigma_larges}.
To estimate the interaction rate, notice that although the average number density of dark radiation is well-approximated by $\langle n_\text{dr} \rangle \simeq \frac{1}{3} R_l n_V$, the distribution of relativistic dark photons is highly inhomogeneous, with $n_\text{dr} \sim \langle n_\text{dr} \rangle H^{-1} / \Delta L \gg \langle n_\text{dr} \rangle$ in a thin shell surrounding the bubble walls and zero elsewhere.
This cross section defines a mean-free-path $\lambda_\text{mfp} = (\sigma n_\text{dr})^{-1}$. Comparing this length scale to the typical thickness of one of these shells we find,~e.g.~
\begin{align}
    \frac{\lambda_\text{mfp}}{\Delta L}
    		%\sim \frac{H(T_*)}{\sigma R_l n_V}
    		\sim \frac{10^{8}}{\lambda_\Phi^2} \left( \frac{m}{1 \ \text{eV}} \right)^3 \left( \frac{10^{-4}}{\Delta m^2 / m^2} \right)^6 \left( \frac{T_*}{100 \ \text{GeV}} \right) \left( \frac{\alpha}{10^{-2}} \right)^2 \left( \frac{\rho_\text{dm}}{\rho_V} \right)^3 \ .
\label{eq:mfpVV}
\end{align}
The above expression has a strong dependence on a number of parameters, especially the fractional change in the mass of the dark photon. But interestingly it can remain $\gg 1$ in a large region of the relevant parameter space where longitudinal reflections are relevant to the evolution of the bubble walls. In this case, the shells of dark radiation will pass each other without significant dissipation, a process that could last for much longer  than the usual duration of a cosmological phase transitions. The long-lasting motion of these shells could greatly enhance the strength of the gravitational wave signal, a possibility that motivates more careful exploration of this potential new source of gravitational waves. We will return to this topic in future work.

\begin{figure}[h]
  \centering
  \includegraphics[scale=0.4]{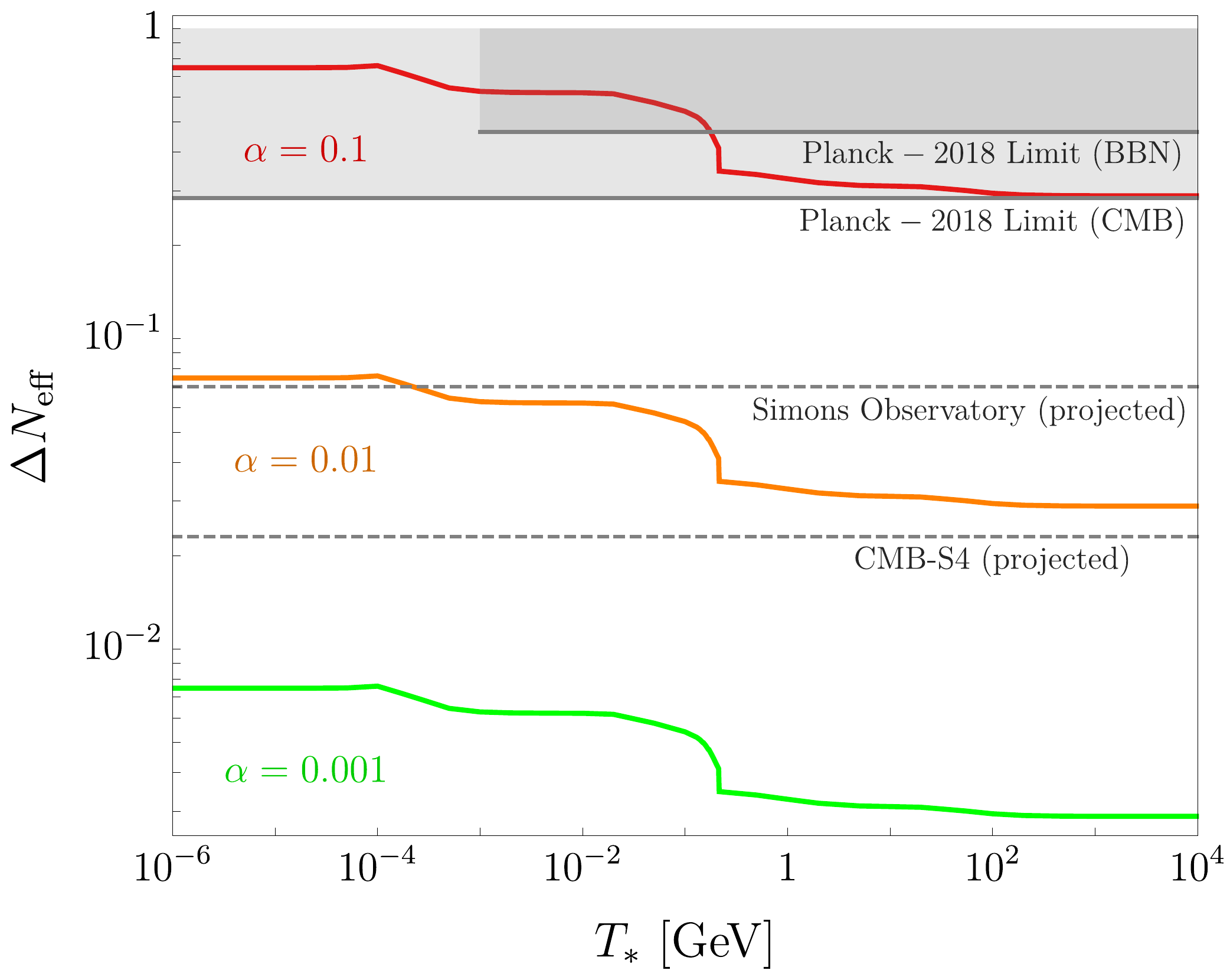}
  \caption{
 In phase transitions where bubble walls reach an equilibrium regime as a result of dark photon reflections, most of the difference in vacuum energy densities goes into turning a fraction of the (cold) dark photons into dark radiation.
  If the dark radiation remains relativistic until late times, an observable contribution to $\Delta N_\text{eff}$ is possible.
  Plotted are the maximum possible contributions to $\Delta N_\text{eff}$, as discussed in the main text, as a function of the thermal plasma at the epoch of the transition, $T_*$. 
  Solid gray lines correspond to current bounds~\cite{Aghanim:2018eyx}, and dashed gray lines show the projected sensitivities of upcoming observatories \cite{CMBstage4,SimonsObservatory:2018koc}.
  }
  \label{fig:DNeffPlot}
\end{figure}

Moreover, it is interesting to consider the limiting possibility that the reflected dark photons remain relativistic long after the phase transition is over, leading to a dark radiation signal.
Making the optimistic assumption that significant losses to gravitational radiation and other ``inelastic" processes can be ignored, the requirement that the reflected dark photons remain relativistic at temperatures $T \leq T_*$ can be written as:
\begin{align}
\label{eq:gammaOfT}
	\gamma_\text{dr} (T \leq T_*)
  		  	\simeq 2 \gamma_\text{eq}^2 \frac{a(T_*)}{a(T)}
			\approx 2 \times 10^{13} \left( \frac{10^{-4}}{\Delta m^2 / m^2} \right)^2 \left( \frac{\alpha}{10^{-2}} \right) \left( \frac{\rho_\text{dm}}{\rho_V} \right) \left( \frac{T}{1 \ \text{MeV}} \right) \gtrsim 1 \ .
\end{align}
Of particular interest are the temperatures of Big Bang Nucleusynthesis ($T_\text{BBN} \sim 1 \ \text{MeV}$) and recombination ($T_\text{rec} \sim 1 \ \text{eV}$), when $\Delta N_\text{eff}$ bounds exist.
As can be inferred from \cref{eq:gammaOfT}, this is easily the case provided $\alpha$ is not too tiny.
When this is the case, the corresponding contribution to $\Delta N_\text{eff}$ can be written as
\begin{align}
    \Delta N_{\rm eff}
    &= \frac{8}{7}\left(\frac{11}{4}\right)^{4/3} \left( \frac{\rho_\text{dr}(T_*)}{\rho_{\gamma , 0}} \right) \left( \frac{a(T_*)}{a (T_0)} \right)^4 \\[2.0ex]
    & \simeq 1.4  \left( \frac{\alpha}{0.1} \right) \frac{g_*(T_*)}{g_{*,s}(T_*)^{4/3}} \\[2.0ex]
    & \simeq 0.3  \left( \frac{\alpha}{0.1} \right) \left( \frac{g_{*}(100 \ \rm GeV)}{g_{*}(T_*)} \right)^{1/3} \ ,
\label{eq:Neff}
\end{align}
where $\rho_{\gamma , 0}$ refers to the current energy density in (Standard Model) photons. In the first step above we have used $\rho_\text{dr} (T_*) \simeq \Delta V$, as discussed around \cref{eq:ndr}-(\ref{eq:rhodr}), and in the final step we have ignored the difference between $g_{*,s}(T_*)$ and $g_{*}(T_*)$.
Interestingly, phase transitions with $\alpha \sim 10^{-1}$ would already be probed by current $\Delta N_\text{eff}$ constraints -- though we emphasize that this is under the assumption that the highly relativistic dark photons suffer no significant energy loss since the time of the phase transition other than redshift due to Hubble expansion.
Under this assumption, \cref{fig:DNeffPlot} shows the prediction for $\Delta N_\text{eff}$ as a function of the characteristic temperature of the phase transition, for various values of $\alpha$.
Strong phase transitions with $\alpha > 0.1$ would be probed by current measurements, while weaker transitions with $\alpha \gtrsim 10^{-2}$ could be probed by CMB-Stage 4 observations~\cite{CMBstage4}.
%%%%%%%%%%%%%%%%%%%%%%%%%%%%%%%%%%%%%%%%%%%%%%%%%%%%%%%%%%%%%%%%%%%%%%%%%%%
%%%%%%%%%%%%%%%%%%%%%%%%%%%%%%%%%%%%%%%%%%%%%%%%%%%%%%%%%%%%%%%%%%%%%%%%%%%
Because bubble acceleration is stalled by the reflection of particles, this effect is a particularly efficient manner of converting the latent heat of the false vacuum to relativistic matter, thus leading to a significant contribution to $\Delta N_{\rm eff}$ if the reflected particles remain relativistic until late times.
When instead bubbles of the new phase are run-away or reach a terminal velocity by dispersing energy into the thermal plasma, only the resulting stochastic background of gravitational waves contributes towards dark radiation, and $\Delta N_\text{eff}$ is a less important observable.

The second scenario highlighted early on in this subsection corresponds to the case where the phase transition sector is in equilibrium with the thermal plasma. In this case, the reflected dark photons will also interact with the population of $\phi$'s in the thermal fluid via \cref{eq:model_EFT_intro} (or its UV-complete version, as described in \cref{sec:model}). Interactions take place at center-of-mass energies $\sqrt{s} / 2 \sim \sqrt{ 2 \, T_* \, \omega_\text{dr} } / 2 \simeq \gamma_\text{eq} \sqrt{m \, T_*}$, and the corresponding cross section is well-approximated by \cref{eq:sigma_lphi_IR} or \cref{eq:sigma_lphi_UV} depending on whether $\sqrt{s}$ falls above or below the cutoff scale of the effective theory. As discussed in \cref{sec:rates}, the corresponding interaction rate can be well above $H(T_*)$, although once more this depends sensitively on the values of the underlying parameters. At any rate, interactions between the reflected dark photons and the thermal plasma provide an additional obstacle for the dark radiation to propagate undisturbed. It is therefore more likely that the energy density in the dark radiation gets distributed within the primordial plasma, in which case the main source of gravitational radiation would come from motion within the thermal fluid, similar to the case of phase transitions that reach an equilibrium regime as a result of thermal pressure.

Determining more precisely under what circumstances the reflected dark photons remain relativistic until late times, and therefore contribute to $\Delta N_\text{eff}$, is an important question worthy of further attention.

%%%%%%%%%%%%%%%%%%%%%%%%%%%%%%%%%%%%%%%%%
\section{\label{sec:conclusions} \ Conclusions}
%%%%%%%%%%%%%%%%%%%%%%%%%%%%%%%%%%%%%%%%%

In this work, we have discussed a new physical effect that can affect the evolution of cosmological vacuum bubbles expanding against a population of phase-dependent massive dark photons, with a special focus on the case where the dark photons furnish the dark matter. Namely, the existence of a transient relativistic regime, for sufficiently thin walls, characterized by a constant reflection probability of longitudinal dark photons.
The reflection of longitudinal modes creates a pressure on the expanding interface that features a characteristic non-monotonic dependence on the $\gamma$-factor of the bubble wall, reaching a peak at intermediate values of $\gamma$ that we have dubbed Maximum Dynamic Pressure. The existence of a MDP that exceeds the asymptotic value of the pressure in the ultra-relativistic limit can make it much harder for bubble walls to become run-away, even in the absence of a thermal plasma that interacts with the wall background.

Our work opens a number of avenues for future exploration.
In phase transitions where bubble walls reach an equilibrium regime as a result of this effect, the later evolution of the reflected dark photons could modify the features of the resulting gravitational wave signal and, in some cases, lead to an observable contribution to $\Delta N_\text{eff}$ if the reflected dark photons remain relativistic until late times. As discussed in \cref{sec:friction} (specially \ref{sec:fate_dr}), the extent to which this happens depends on a variety of considerations, most importantly on whether the sector undergoing vacuum decay is cold or, instead, is in thermal equilibrium with the primordial plasma surrounding the expanding bubbles.
Understanding more generally the possible implications of the dark radiation for these two broad classes of models is clearly an important topic that deserves further attention.

Moreover, although we have focused exclusively on the case of phase-dependent massive dark photons, it is possible that the phenomenon of MDP on expanding bubble walls could be realized in scenarios beyond this example.
Given the impact this can have on the dynamics of bubble walls, which in turn largely determine the features of the resulting gravitational wave signal, this is an important question worthy of further investigation.

%%%%%%%%%%%%%%%%%%%%%%%%%%%%%%%%%%%%%%%%%
\section*{\label{sec:acknowledgments} \ Acknowledgments}
%%%%%%%%%%%%%%%%%%%%%%%%%%%%%%%%%%%%%%%%%
We thank Alex Azatov, Nathaniel Craig, Yann Gouttenoire, John March-Russell, Mehrdad Mirbabayi, Ken Van Tilburg and Govanni Villadoro for helpful discussions.
IGG gratefully acknowledges support from NSF Grant PHY-2207584, and from the James Arthur Postdoctoral Fellowship at NYU.
RPB is grateful for the support and hospitality of the KITP through its graduate fellowship program, which resulted in the present collaboration. At the KITP, this research was supported in part by the National Science Foundation under Grant No.~NSF PHY-1748958.
The research of GK was supported by the Len DeBenedictis Graduate Fellowship and DOE grant number DE-SC0011702. 

%%%%%%%%%%%%%%%%%%%%
%%%%%%%%%%%%%%%%%%%%
\appendix

%%%%%%%%%%%%%%%%%%%%%%%%%%%%%%%%%%%%%%%%%%%%%%%%%%%%%%%%%%%%
\section{Beyond the effective theory}
%%%%%%%%%%%%%%%%%%%%%%%%%%%%%%%%%%%%%%%%%%%%%%%%%%%%%%%%%%%%

\subsection{Abelian Higgs UV-completion}
%%%%%%%%%%%%%%%%%%%%%%%%%%%%%%%%%%%%%%%%%%%%%%%%%%%%%%%%%%%%
\label{sec:model}

In this appendix, we briefly discuss how the effective Lagrangian of \cref{eq:L_dim4}-(\ref{eq:model_EFT_intro}) may arise from a more complete framework. Our goal is not to be exhaustive, but rather to identify the necessary features of underlying models giving rise to our effective theory.
The simplest UV-completion of \cref{eq:L_dim4}-(\ref{eq:model_EFT_intro}) can be written in terms of an Abelian Higgs model:
\begin{equation}
	\mathcal{L}_\text{UV} = - \frac{1}{4} F_{\mu \nu} F^{\mu \nu} + | D_\mu \Phi |^2 - V_\Phi (|\Phi|) + \frac{1}{2} (\partial_\mu \phi)^2 - V_\phi (\phi) \, + \, \frac{\eta}{2} \phi^2 |\Phi|^2 \ ,
\label{eq:model_UV}
\end{equation}
where $D_\mu \Phi \equiv (\partial_\mu - i g' V_\mu ) \Phi$. Assuming a Higgs potential of the usual form: $V_\Phi (|\Phi|) = \lambda_\Phi / 2 (|\Phi^2| - v'^2/2)^2$, the corresponding radial mode gets a mass $m_\rho = \sqrt{\lambda_\Phi} v'$, whereas $m = g' v'$ is the mass of the dark photon.
Integrating out the radial mode, one obtains a low energy effective theory featuring additional interactions among the light degrees of freedom. In particular, a term such as \cref{eq:model_EFT_intro} -- repeated here for convenience -- is generated at tree-level:
\begin{equation}
\label{eq:kappa_abstract}
   	 \mathcal{L} \supset \frac{\kappa}{2} \phi^2 V^\mu V_\mu
 	\qquad \quad \text{with} \qquad \quad
	\kappa =  \frac{\eta \, m^2}{m_\rho^2} = \frac{\eta g'^2}{\lambda_\Phi} \ .
\end{equation}
Notice in particular that the ratio $\Delta m^2 / m^2$ is independent of $g'$, and it is therefore controlled by a different set of parameters of the underlying model compared to the overall mass of the dark photon.
In this context, the scale of UV-completion is $\Lambda = m_\rho$, which can be written as
\begin{equation}
	m_\rho = \frac{\sqrt{\eta} v}{\sqrt{\Delta m^2 / m^2}} \lesssim \frac{4 \pi v}{\sqrt{\Delta m^2 / m^2}} \ .
\label{eq:cutoffUV}
\end{equation}
This reproduces our EFT expectation for the upper bound on the cutoff scale in \cref{eq:cutoff_intro} subject to the perturbativity requirement $\eta \lesssim 16 \pi^2$.

Of course, the low energy effective theory contains interactions among the light degrees of freedom beyond \cref{eq:model_EFT_intro}.
Of particular interest are self-interactions among the dark photons. Indeed, a quartic interaction is generated at tree-level, of the form
\begin{equation}
	\mathcal{L} \supset \lambda_V (V_\mu V^\mu)^2
	\qquad \quad \text{with} \qquad \quad \lambda_V \sim \frac{g'^4}{\lambda_\Phi} = \lambda_\Phi \frac{m^4}{m_\rho^4} \lll 1 \ .
\label{eq:lambdav}
\end{equation}

Interactions among the dark photons, as well as between the dark photons and a potential population of $\phi$ particles in the thermal plasma, are relevant (\emph{i}) to assess the extent to which our assumptions that the dark matter is cold and non-interacting are self-consistent, and (\emph{ii}) to determine the evolution of the reflected dark photons, which is the topic of \cref{sec:fate_dr}.
For convenience, we summarize the relevant interaction rates in the remainder of this appendix.

\subsection{Interaction rates}
%%%%%%%%%%%%%%%%%%%%%%%%%%%%%%%%%%%%%%%%%%%%%%%%%%%%%%%%%%%%
\label{sec:rates}

At low momenta, \cref{eq:lambdav} leads to a self-interaction among the dark photons of the form
\begin{equation}
  \sigma_{\scriptstyle VV} \sim \frac{\lambda_V^2}{8 \pi m^2} = \frac{\lambda_\Phi^2}{8 \pi m_\rho^2} \left( \frac{m}{m_\rho} \right)^6
  \qquad \text{for} \qquad |\vec k| \ll m  \ .
  \label{eq:sigma_smallk}
\end{equation}
The corresponding dark matter self-interaction rate is always tiny compared to the Hubble rate in the early Universe in the region of parameter space relevant to this work.
Potentially more relevant are interactions between the dark photons and $\phi$ particles in the thermal fluid. In particular, processes of the form $V \phi \rightarrow V \phi$ could be efficient at transferring energy from the plasma to the dark matter, unless the corresponding interaction rates are tiny.
The relevant cross section is of the form, parametrically:
\begin{equation}
	\sigma_{\scriptstyle V\phi} \sim \frac{\kappa^2}{8 \pi T_*^2} \sim \frac{1}{8 \pi T_*^2} \left( \frac{\Delta m^2}{m^2} \right)^2 \frac{m^4}{T_*^4} \ ,
\end{equation}
where $\kappa \simeq \Delta m^2 / v^2$, as discussed below \cref{eq:model_EFT_intro}, and we've taken $m_\phi \sim v \sim T_*$, which should be a good approximation at the epoch of the phase transition (barring a tiny quartic coupling for $\phi$).
Assuming a thermal population of $\phi$ particles, with number density $n_\phi \sim T_*^3$, we find
\begin{equation}
	\frac{\Gamma_{\scriptstyle V\phi}}{H(T_*)} \sim 10^{-39} \left( \frac{\Delta m^2 / m^2}{10^{-4}} \right)^2 \left( \frac{m}{1 \ \text{eV}}\right)^4 \left( \frac{100 \ \text{GeV}}{T_*}\right)^5 \ .
\end{equation}
Clearly, this ratio can be very small in much of the parameter space of interest (see left panel in \cref{fig:models}).

Let us know discuss interactions involving the (highly relativistic) reflected dark photons.
These will be relevant in \cref{sec:fate_dr} when we discuss the fate of the dark radiation, and for convenience we summarize the relevant results here.
At large center-of-mass energies, i.e.~$\sqrt{s} \gg m$, the two-to-two scattering cross section among the dark photons is dominated by the scattering of longitudinally-polarized vectors, and so it will be sufficient for our purposes to just consider the process $V_l V_l \rightarrow V_l V_l$.
Two limiting kinematic regimes are of interest.
In the region $m \ll \sqrt{s} \ll m_\rho$, the relevant cross section is of the form
\begin{equation}
	\sigma_{\scriptstyle ll} \sim \frac{\lambda_V^2}{8 \pi s} \left( \frac{\sqrt{s}}{m} \right)^8
				\sim \frac{\lambda_\Phi^2}{8 \pi m_\rho^2} \left( \frac{\sqrt{s}}{m_\rho} \right)^6
				\qquad \text{for} \qquad
				m  \ll \sqrt{s} \ll m_\rho \ .
	\label{eq:sigma_smalls}
\end{equation}
This will be relevant for scattering between the dark radiation and the cold dark photons, where the relevant center-mass-energy is given by $\sqrt{s}/2 \simeq \gamma_\text{eq} m$. Comparing the relevant interaction rate to $H(T_*)$, we find
\begin{equation}
	\frac{\Gamma_{\scriptstyle ll}}{H (T_*)} \sim 10^{-12} \, \frac{\lambda_\Phi^2}{\eta^4} \left( \frac{m}{1 \ \text{eV}} \right)^5 \left( \frac{10^{-4}}{\Delta m^2 / m^2} \right)^2 \left( \frac{T_*}{v} \right)^8 \left( \frac{100 \ \text{GeV}}{T_*} \right)^4
	\left( \frac{\alpha}{10^{-2}} \right)^3 \left( \frac{\rho_\text{dm}}{\rho_V} \right)^2 \ ,
\end{equation}
where we have evaluated the cross section at $\sqrt{s}/2 = \gamma_\text{eq} m$, and substituted $m_\rho$ as given in \cref{eq:cutoffUV}.
Notice that the ratio $T_* / v$ could be $\lll 1$ if the phase transition sector is cold and decoupled from the thermal plasma, as discussed below \cref{eq:L_dim4}, which would further suppress the above ratio.
If instead $\sqrt{s} \gg m_\rho$ then the cross section must be computed in the complete theory of \cref{eq:model_UV}, and we find
\begin{equation}
	\sigma_{\scriptstyle ll}	\sim \frac{\lambda_\Phi^2}{8 \pi s}
			\qquad \text{for} \qquad \sqrt{s} \gg m_\rho \ ,
	\label{eq:sigma_larges}
\end{equation}
as expected by virtue of Goldstone Equivalence. This is the form of the cross section used in deriving \cref{eq:mfpVV}.

Another important class of interactions are those between the dark radiation and the $\phi$ particles in the plasma when the phase transition sector is thermal.
In particular, if the process $V_l \phi \rightarrow V \phi$ happens efficiently, the energy density stored in the dark radiation would quickly become distributed among the thermal fluid, as discussed in \cref{sec:fate_dr}.
The typical $s$-parameter is of the form $s \sim 2 T_* \omega_\text{dr}$.
In the regime $\sqrt{s} \ll m_\rho$,
\begin{equation}
\label{eq:sigma_lphi_IR}
	\sigma_{\scriptstyle l\phi} \sim \frac{\kappa^2}{8 \pi s}\left( \frac{\sqrt{s}}{m} \right)^4
  \qquad \text{for} \qquad m  \ll \sqrt{s} \ll m_\rho \ ,
\end{equation}
whereas in the high-energy limit $\sqrt{s} \gg m_\rho$,
\begin{align}
\label{eq:sigma_lphi_UV}
   \sigma_{\scriptstyle l\phi} &\sim \frac{\eta^2}{8 \pi s} 
     \qquad \text{for} \qquad \sqrt{s} \gg m_\rho \ .
\end{align}
As an example, if the there's a full thermal distribution of $\phi$ particles with $n_\phi \sim T_*^3$, then the relevant interaction rate, relative to $H(T_*)$ is given by
\begin{equation}
	\frac{\Gamma_{\scriptstyle l\phi}}{H(T_*)} \sim 10^8 \, \eta^2 \left( \frac{\Delta m^2}{m^2} \right)^2 \left( \frac{1 \ \text{eV}}{m} \right)^2 \left( \frac{100 \ \text{GeV}}{T_*} \right)^2 \left( \frac{10^{-2}}{\alpha} \right)^2 \left( \frac{\rho_V}{\rho_\text{dm}} \right)^2 \ ,
\end{equation}
where we have used \cref{eq:sigma_lphi_UV} in obtaining $\Gamma_{\scriptstyle l\phi}$, as appropriate given our choice of parameters on the right-hand-side. Clearly, interactions between the dark radiation and the thermal plasma can be very efficient in parts of the relevant parameter space but -- again -- this conclusion depends strongly on the choices of the various scales and couplings.

%%%%%%%%%%%%%%%%%%%%%%%%%%%%%%%%%%%%%%%%%%%%%%%%%%%%%%%%%%%%
\section{Reflection probabilities -- supplemental material}
%%%%%%%%%%%%%%%%%%%%%%%%%%%%%%%%%%%%%%%%%%%%%%%%%%%%%%%%%%%%

This appendix contains additional details supplementing the discussion of \ref{sec:probs}.

%%%%%%%%%%%%%%%%%%%%%%%%%%%%%%%%%%%%%%%%%%%%%%%%%%%%

\subsection{Longitudinal reflection from Goldstone Equivalence}
\label{sec:get}

In the main text we described the theory of a massive vector in \cref{eq:L_dim4,eq:EOM} in terms of the Proca Lagrangian, featuring three degrees of freedom for the three physical polarizations of a massive spin-1 particle.
However, this is often understood as a particular choice of gauge, after gauge redundancy is restored via $V_\mu \longrightarrow V_\mu - \partial_\mu \theta$. Although $\theta=0$ (unitary gauge) is naively simplest, often it is convenient to keep the `Goldstone' $\theta$.
The Goldstone Equivalence Theorem states that in the relativistic limit $\omega \gg m $ correct amplitudes can be obtained by identifying the physical longitudinal vector with the scalar $\theta$, with an error of order $\mathcal{O}\left(m^2/\omega^2\right)$. 
Using the usual gauge-fixing parameter $\xi$, the theory is described by
\begin{equation}
  \mathcal{L} = -\frac{1}{4} F^{\mu\nu}F_{\mu\nu} + \frac{1}{2}m_V^2(z) (V_\mu - \partial_\mu \theta)^2 - \frac{1}{2\xi}\left( \partial_\mu V^\mu + m^2 \xi \theta \right)^2 \ .
 \label{eq:ProcaRxi}
\end{equation}
Proceeding with the equation of motion for $\theta$:
\begin{equation}
  \partial_\mu\left( m_V^2(z) \partial^\mu \theta \right) + \xi m^4 \theta=0 \label{eq:GETeom} ~,
\end{equation}
where we have neglected mixing with $V^\mu$ in order to test the spirit of the Goldstone Equivalent in our context. Working in the step function limit, the scattering solution is simple for $z\neq 0$:
\begin{equation}
  \theta = e^{- i \omega t }\left\{ \begin{matrix} & e^{ik_z} + r\ e^{-ik_z}, \quad &z<0 \\ & t \ e^{i\tilde{k}_z} , \quad &z>0 \end{matrix}\right., 
\end{equation}
\begin{equation}
    \quad \quad k_z = \sqrt{\omega^2-\xi m^2}~, \quad \tilde{k}_z = \sqrt{\omega^2-\xi m^4/\tilde{m}^2} ~,
\end{equation}
and need only be supplemented by the matching conditions:
\begin{equation}
    m_V^2(z) \partial_z\theta  \qquad \text{and} \qquad \theta \qquad \text{are continuous at } z=0 \ ,
\end{equation}
derived by integrating \cref{eq:GETeom}  once and twice respectively. A little algebra gives the reflection probability as
\begin{equation}
    R = |r|^2 =  \left( \frac{\tilde{k}_z \tilde{m}^2 - k_z m^2}{\tilde{k}_z \tilde{m}^2 + k_z m^2}\right)^2 \xrightarrow{\omega \gg m, \tilde{m}} \left( \frac{\tilde{m}^2 - m^2}{\tilde{m}^2 + m^2}\right)^2 + \mathcal{O}\left(\tilde{m}^2/\omega^2\right) \,
\end{equation}
which is $\xi$ (gauge) independent and matches the leading order result derived in the main text \cref{eq:RL} consistent with the spirit of the GET.  
This result suggests that the enhanced reflection of the inter-relativistic limit described in this work might fundamentally be a property of Nambu-Goldstone bosons more generally, a topic that we will return to in future work.

%%%%%%%%%%%%%%%%%%%%%%%%%%%%%%%%%%%%%%%%%%%%%%%%%%%%%%%%%%%%
\subsection{Scattering on a $\delta'$ potential}
%%%%%%%%%%%%%%%%%%%%%%%%%%%%%%%%%%%%%%%%%%%%%%%%%%%%%%%%%%%%
\label{app:deltaprime}

As anticipated in \ref{sec:smooth}, the effective scattering potential for the longitudinal component in the limit of vanishing wall thickness takes the form
\begin{equation}
	U_l(z) \xrightarrow{L \rightarrow 0} \kappa \, \delta'(z) \qquad \text{with} \qquad \kappa \equiv - \frac{\Delta m^2}{2 m^2} \ ,
\end{equation}
and \cref{eq:full_long} reads
\begin{equation}
	(\partial_z^2 + k_z^2) \lambda(z) = \kappa \, \delta'(z) \lambda(z) \ .
\label{eq:Sch_deltap}
\end{equation}
We can obtain matching conditions for $\lambda$ and $\lambda'$ by integrating (twice) over this equation.
First, integrating \cref{eq:Sch_deltap} from $z_0 < 0$ to $z$, we find
\begin{gather} 
	\int_{z_0}^z d\hat z \, \left[ (\partial_{\hat z}^2 + k_z^2) \lambda(\hat z) \right] = \int_{z_0}^z d\hat z \, \kappa \, \delta'(\hat z) \lambda(\hat z) \\
	\Rightarrow \qquad \lambda'(z) - \lambda'(z_0) + k_z^2 \int_{z_0}^z d\hat z \, \lambda(\hat z) = \kappa \, \lambda(z) \delta(z) - \frac{\kappa}{2} \left[ \lambda' (0^+) + \lambda' (0^-)  \right] \Theta(z) \ .
	\label{eq:Sch_deltap_int}
\end{gather}
Integrating \cref{eq:Sch_deltap_int} from $z= - \epsilon$ to $z= + \epsilon$ and taking the limit $\epsilon \rightarrow 0$, we find:
\begin{equation}
	\lambda(0^+) - \lambda(0^-) = - \frac{\kappa}{2} \left[ \lambda(0^+) + \lambda(0^-) \right] \ .
\end{equation}
Similarly, taking $z_0 = - \epsilon$ and $z = + \epsilon$ in \cref{eq:Sch_deltap_int}, with $\epsilon \rightarrow 0$, one finds:
\begin{equation}
	\lambda'(0^+) - \lambda'(0^-) \simeq \frac{\kappa}{2} \left[ \lambda'(0^+) + \lambda'(0^-) \right] \ ,
\end{equation}
where in the right-hand-side we have only kept terms of $\mathcal{O} (\kappa)$.
From the last two equations, one finds $r_l \simeq \kappa$, and therefore
\begin{equation}
	R_l \simeq \kappa^2 = \left( \frac{\Delta m^2}{2 m^2} \right)^2 \ , 
\end{equation}
in agreement with \cref{eq:RL} at leading order in $\kappa \ll 1 $.

%%%%%%%%%%%%%%%%%%%%%%%%%%%%%%%%%%%%%%%%%%%%%%%%%%%%%%%%%%%%
\subsection{The Born approximation}
%%%%%%%%%%%%%%%%%%%%%%%%%%%%%%%%%%%%%%%%%%%%%%%%%%%%%%%%%%%%
\label{app:born}

In \cref{sec:smooth}, we found it convenient to recast the equations of motion for both transverse and longitudinal components as a one-dimensional Schr\"odinger equation, of the form
\begin{equation}
    (\partial_z^2 + k_z^2) \psi (z) = U(z) \psi (z) \ ,
\label{eq:1dqm}
\end{equation}
with $U (z) \rightarrow U_l (z)$ and $\psi \rightarrow \xi$ for the longitudinal component (c.f.~\cref{eq:full_long}), and $U (z) \rightarrow U_\perp (z)$ and $\psi \rightarrow v^\mu_\perp$ for the transverse modes (c.f.~\cref{eq:full_perp}). It is an important feature that the effective scattering potential vanishses in the limit $z \rightarrow - \infty$, as will become clear in due time.

As is well-known, it is possible to recast \cref{eq:1dqm} as an integral equation, as follows
\begin{equation}
    \psi (z) = \psi_0 (z) + \int_{-\infty}^{\infty} dz' \ G(z-z') U(z') \psi (z') \ ,
    \label{eq:born_integral_eqn}
\end{equation}
where $\psi_0$ is any function satisfying the free particle equation, $\big( \partial_z^2 + k_z^2 \big) \psi_0 = 0$, and $G(z)$ is a Green's function for the differential operator on the right-hand-side of \cref{eq:1dqm}, i.e.~
\begin{equation}
    \big( \partial_z^2 + k_z^2 \big) G (z) = \delta (z) \ .
    \label{eq:born_green}
\end{equation}
It is straightforward to check that \cref{eq:born_integral_eqn} is equivalent to \cref{eq:1dqm} by applying $(\partial_z^2 + k_z^2)$ to both sides and using \cref{eq:born_green}.
In order to solve for $G(z)$, note that for $z \neq 0$ \cref{eq:born_green} is just the free particle equation. The matching conditions at $z=0$ are that $G$ must be continuous and $G'$ must have unit jump. Putting this together, the solutions are
\begin{equation}
    G(z) = \pm \frac{1}{2ik_z} e^{\pm ik_z |z|} \ ,
\end{equation}
up to addition of functions that satisfy the free particle equation. It turns out that we only need one of these two solutions: the one with the $+$ sign.

To represent an incoming plane wave, we take $\psi_0 (z) = e^{ik_z z}$. Then our integral equation is
\begin{equation}
    \psi (z) = e^{ik_z z} + \frac{1}{2ik_z} \int_{-\infty}^{\infty} dz' \ e^{ik_z |z-z'|} \ U(z') \psi (z') \ .
\end{equation}
So far, we have made no approximations. However, if the correction to the incoming wavefunction $\psi_0$ is small (meaning the reflection coefficient is tiny), then we can plug this equation for $\psi$ into itself in the integral and truncate higher order terms, leaving
\begin{align}
    \psi (z) &\simeq e^{ik_z z} + \frac{1}{2ik_z} \int_{-\infty}^{\infty} dz' \ e^{ik_z |z-z'|} \ U(z') e^{ik_z z'} \\
    &= e^{ik_z z} + \frac{e^{ik_z z}}{2ik_z} \int_{-\infty}^{z} dz' \ U(z') + \frac{e^{-ik_z z}}{2ik_z} \int_{z}^{\infty} dz' \ e^{2ik_z z'} \ U(z') \ .
\end{align}
We wish to extract a reflection probability from this `first Born approximation'. To this end, consider the limit $z\to -\infty$. The second term above clearly vanishes, given that $U(z') \to 0$ as $z' \to -\infty$ as emphasized below \cref{eq:1dqm}. Furthermore, the third term can be identified as the reflected plane wave piece, with associated probability
\begin{equation}
    R_{l\text{,\,Born}} = \frac{1}{4k_z^2} \left| \int_{-\infty}^{\infty} dz \ e^{2ik_z z} \ U(z) \right|^2 \ .
    % \label{eq:born_prob}
\end{equation}
Keep in mind that $R_l \simeq R_{l\text{,\,Born}}$ only when $R_l \ll 1$.

%%%%%%%%%%%%%%%%%%%%%%%%%%%%%%%%%%%%%%%%%%%%%%%%%%%%%%%%%%%%
\subsection{Numerical methods}
%%%%%%%%%%%%%%%%%%%%%%%%%%%%%%%%%%%%%%%%%%%%%%%%%%%%%%%%%%%%
\label{app:numerics}

Finding the reflection and transmission coefficients for the longitudinal mode reduces to solving \cref{eq:full_long} with boundary conditions corresponding to plane-wave behavior far away from the bubble wall. i.e.~
\begin{equation}
	\lambda (z) = \begin{cases} e^{i k_z z} + r_l e^{- i k_z z} 	& \text{as} \quad z \rightarrow - \infty \\
					t_l e^{i \tilde{k}_z z} 					& \text{as} \quad z \rightarrow + \infty
					\end{cases} 
\label{eq:xi}
\end{equation}
It will be convenient to rescale $\lambda$ by an overall complex constant, and define $f(z) \equiv t_l^{-1} \lambda(z)$. The equation of motion for $f(z)$ remains as in Eq.(\ref{eq:full_long}), and the boundary conditions can now be written as $f (z_0) = 1$ and $\partial_z f (z_0) = i \tilde{k}_z$ for some $z_0 > 0$ far away to the right of the wall. Our numerical solution will lie in the window $z \in [-z_0, z_0]$, where $z_0$ is set to be much larger than the incident particle wavelength (constraining for low energies) and the thickness of the bubble wall (constraining for high energies).

Lastly, we need a prescription for extracting the reflection probability from the numerical solution. Consider evaluating the asymptotic expression for $f(z)$ at the following points, where $n \in \mathbb{N}$, $n \to \infty$:
\begin{align}
    f \left( -\frac{n\pi}{k_z} \right) &\longrightarrow \left( t_l^{-1} + t_l^{-1} \, r_l \right) (-1)^n \\
    f \left( -\frac{(n-\frac{1}{2})\pi}{k_z} \right) &\longrightarrow \left( t_l^{-1} - t_l^{-1} \, r_l \right) (-1)^n (i)
\end{align}
These equations can then be easily inverted to obtain the scattering coefficients $r_l$ and $t_l$, and ultimately the reflection and transmission probabilities.

%\pagebreak

\bibliography{refs_bubble_walls}
\bibliographystyle{utphys}

\end{document}